\documentclass{aastex}
\usepackage{natbib,epsfig,rotate}
\usepackage[onecolumn]{emulateapj5}

\def\be{\begin{eqnarray}}
\def\ee{\end{eqnarray}}
\def\half{\frac{1}{2}}
\def\etal{{\it et al.$\,\,$}}
\def\simless{\mathbin{\lower 3pt\hbox
   {$\rlap{\raise 5pt\hbox{$\char'074$}}\mathchar"7218$}}} %< or of order
\def\simgreat{\mathbin{\lower 3pt\hbox
   {$\rlap{\raise 5pt\hbox{$\char'076$}}\mathchar"7218$}}} %> or of order

\def\vperp{{V_{\perp}}}

\def\vperpvec{{\bf V_{\perp}}}

\def\vpperpvec{{\bf   {V_{\rm p}}_\perp}}
\def\vobsperpvec{{\bf {V_{\rm obs}}_\perp}}
\def\vismperpvec{{\bf {V_{\rm screen}}_\perp}}

\def\thd{{\theta_{\rm d}}}

\def\thvec{{ \mbox{\boldmath $\theta$} }}

\def\psivec{{ \mbox{\boldmath $\psi$} }}
\def\kappavec{{ \mbox{\boldmath $\kappa$} }}

\def\fthvec{{\mbox{$B$}}}    % pdf of vector
\def\ds{{D_{\rm s}}}        
\def\dds{{D - D_{\rm s}}}   

\def\cm3{{cm$^{-3}$}}

\def\vperpvechat{{\mbox{\boldmath $\hat V_{\perp}$}}}

\def\rf{{r_{\rm F}}}
\def\mbsq{{m_{\rm B}^2}}
\newcommand{\fnu}{f_{\nu}}

\newcommand{\ft}{f_{\rm t}}

\newcommand{\linner}{\ell_1}
\newcommand{\sfphi}{D_{\phi}}
\newcommand{\dt}{D -\ds}

\def\bvec{{ \mbox{\boldmath $b$} }}
\newcommand{\Cmatrix}{\bf C}
\newcommand{\fnuab}{f_{\nu}}
\newcommand{\ftab} {f_{ {\rm t}}}

\newcommand{\thvecbar}{{ \mbox{$\overline\thvec$} }}

\def\nubar{{\nu_0}}
\newcommand{\bfu}{{ \mbox{\boldmath $u$} }}
\newcommand{\thvecp}{{ \mbox{\boldmath $\theta_p$} }}

\newcommand{\rvec}{{\mbox{\boldmath  $r$} }}
\newcommand{\rsvec}{{\mbox{\boldmath  $r_s$} }}
\newcommand{\rvecp}{{{\bf r}^{\prime}}}
\def\phiij{{\Phi}}

\def\dse{{D_{\rm e}}}
\def\fpq{{S_2}}
\newcommand{\pbar}{{\overline p}}
\newcommand{\qbar}{{\overline q}}
\newcommand{\taud}{\tau_{\rm d}}	% pulse broadening time
\newcommand{\dtd}{\Delta t_{\rm d}}	% ISS time scale

\newcommand\qabs{\vert q \vert}

\begin{document} 
\title{ Theory of Parabolic Arcs in Interstellar Scintillation 
	Spectra }
\author{	James M. Cordes\altaffilmark{1},
		Barney J. Rickett\altaffilmark{2}, 
		Daniel R. Stinebring\altaffilmark{3} and
		William A. Coles\altaffilmark{2} 
		\\ 
		~\\
		%Draft \today
		%22 June 2004
}
\altaffiltext{1}{Department of Astronomy, Cornell University, Ithaca, NY 14853; cordes@astro.cornell.edu}
\altaffiltext{2}{Department of Electrical Engineering and Computer Science, UC San Diego, La Jolla, CA 92093; rickett@ece.ucsd.edu, coles@ece.ucsd.edu} 
\altaffiltext{3}{Department of Physics and Astronomy, Oberlin College, Oberlin, OH 44074; Dan.Stinebring@oberlin.edu}

\begin{abstract}
Interstellar scintillation (ISS) of pulsar emission
is caused by small scale ($\ll 1$ AU) structure in the ionized 
interstellar medium (ISM) and appears as modulations
in the dynamic spectrum, radio intensity versus time and frequency.
We develop a theory that relates the two-dimensional power spectrum of 
the dynamic spectrum, the secondary spectrum,  
to the scattered image of a pulsar emanating from a thin scattering
screen.
Recent work has identified parabolic-shaped arcs in secondary spectra
for many pulsars.  We show that arcs are generic features
to be expected from media that scatter radiation at angles much larger than
the root-mean-square scattering angle.   Each point in the secondary spectrum
corresponds to a sinusoidal fringe pattern in the dynamic spectrum whose
periods in time and frequency
can be related to the differences in arrival time delay
and in fringe rate (or Doppler frequency) between pairs of 
components in the scattered image.  
The arcs correspond to a parabolic 
relation between the periods through their 
common dependence on the angle of arrival of scattered components.
Arcs appear even without consideration of the dispersive nature of the
plasma and are particularly simple to analyze in the weak scintillation regime,
in which case the secondary spectrum can be inverted to estimate the phase
perturbation in the screen.
Arcs are more prominent in media with negligible inner scale
and with wavenumber spectra that are shallower than a
(wavenumber)$^{-4}$ spectrum, including the Kolmogorov spectrum
with index $11/3$.
Arcs are also  enhanced when the scattered image is
elongated along the velocity direction, making
them useful for probing  anisotropic structure in the ISM.
The arc phenomenon can be used, therefore, to determine or place limits
on the inner scale and on the anisotropy of scattering irregularities
for lines of sight to relatively nearby pulsars.
Arcs are truncated by finite source size  and
thus provide high angular resolution  ($\lesssim 10^{-6}$ arc sec)
for probing emission regions in pulsars and
compact active galactic nuclei.  Overall, we expect arcs from a given
source to be more prominent at higher frequencies.  
Additional arc phenomena seen from
some objects at some epochs include the appearance of multiple arcs, 
most likely from two or 
more discrete scattering screens along the propagation path, and
arc substructure consisting of small arclets oriented oppositely to
the main arc that persist for long durations and indicate
the occurrence of long-term multiple images from the scattering screen.
\end{abstract}

\keywords{ISM: general --- ISM: structure --- 
pulsars: general --- scattering}

\section{ Introduction}

Pulsar scintillations have been used to study the small-scale
structure in the electron density in the interstellar medium (ISM).
This microstructure covers a range of scales extending from 
$\lesssim 10^3$ km to at least 10 AU following a Kolmogorov-like
wavenumber spectrum. That is,  the slope of the three dimensional
wavenumber spectrum is often estimated to be roughly $-11/3$,
though there is also evidence that some lines of sight
at some epochs show excess power in the wavenumber spectrum on scales
near 1 AU.   The chief observable phenomena are intensity fluctuations 
in time and frequency (interstellar scintillation or ISS), 
angular broadening (`seeing'), pulse broadening, 
and arrival time variations.  These effects
have been well studied in the three decades 
since pulsars were discovered. Although refractive  ISS (RISS) has also been 
recognized in compact extra-galactic radio sources,
their larger angular sizes suppress the fine diffractive
scintillations (DISS) which probe the smallest scales in the medium.

Observers have long used dynamic spectra of pulsars to
study the diffractive interstellar scintillation intensity versus time 
and frequency, $S(\nu, t)$. Its two-dimensional correlation function 
is typically used to estimate the characteristic bandwidth 
and time scale for the scintillations.  Frequency dependent 
refraction by large scale irregularities can also modulate these 
diffractive quantities (see e.g. Bhat et al. 1999).  Some observers
noted occasional periodic ``fringes'' in the dynamic spectrum which
they quantified using a two-dimensional Fourier analysis of the 
dynamic spectrum (Ewing et al. 1970; Roberts \& Ables 1982; 
Hewish, Wolszczan, \& Graham 1985; Cordes \& Wolszczan 1986; 
Wolszczan \& Cordes 1987; 
and Rickett, Lyne \& Gupta 1997).
We now refer to the two-dimensional power
spectrum of $S(\nu, t)$  
%(i.e.\ the Fourier transform of the
%correlation function) 
as the ``secondary'' spectrum
of the scintillations.  Fringes in the dynamic spectrum appear as discrete
features in the secondary spectrum and are typically explained as 
interference between two or more scattered images.  
%A related study by Armstrong \& Rickett (1981) examined
%a one-dimensional transform related to the auto-correlation
%of the scattered pulse shape; this function depends
%on the wavenumber spectrum of density irregularities
%and provided some of the evidence for a Kolmogorov spectrum.
Recent observations of secondary spectra with much 
higher dynamic range have identified the dramatic 
``parabolic arc'' phenomenon (Stinebring et al. 2001; hereafter Paper 1).    
Arcs appear as enhanced power (at very low levels) 
along parabolic curves extending out from the origin well 
beyond the normal diffractive feature. 
%bjr  I removed text that started to describe the arc phenomenon
% in more detail, since it is done in the next section.
%
% when the secondary spectrum is displayed at 
% amplitude levels $\lesssim 10^{-3}$ from the peak.
%*bjr Once found, scintillation arcs can be identified in the 
%*bjr published literature going back two decades.
In this paper, we develop a theory 
that relates the secondary spectrum to the underlying 
scattered image of the pulsar.  We find that the arc 
phenomenon is a generic feature of forward scattering 
and we identify conditions that enhance or diminish arcs.
This paper contains an explanation for arcs referred to in Paper 1
and also in a recent paper by Walker et al. (2004), who present
a study of the arc phenomenon using approaches that complement ours.
In \S\ref{sec:obs} we review the salient 
observed features of scintillation arcs. Then in
\S\ref{sec:theory} we introduce the general theory of secondary spectra
through its relation to angular scattering and provide examples that
lead to generalizations of the arc phenomenon.
In \S\ref{sec:ismcases} we discuss cases relevant to the interpretation
of observed phenomena. In \S\ref{sec:scattphys} 
we discuss the scattering physics and show examples of parabolic arcs
from a full screen simulation of the scattering, and we end
with discussion and conclusions in \S\ref{sec:discuss}.

%\input arctheory.v11.sec2.tex

% Dan's changes as of 2004-06-04
% some additional small changes on 2004-06-15  DRS
% These include comments from our telecon on Friday, 2004-06-11.
% Some small changes 2004/06/16 JMC
\section{Observed Phenomena}
\label{sec:obs}

The continuum spectrum emitted by the pulsar
is deeply modulated by interference associated with 
propagation through the irregular, ionized ISM.
Propagation paths change with time owing to motions of the observer,
medium and pulsar, 
causing the modulated spectrum to vary.  
The dynamic spectrum, $S(\nu, t)$, 
is the main observable quantity for our study.
It is obtained by summing over the on-pulse portions of several to 
many pulse periods in a multi-channel spectrometer covering
a total bandwidth up to $\sim 100$  MHz, for durations $\sim 1$ hr.   
We compute its two-dimensional power spectrum 
$S_2(\fnu,\ft) = | \tilde S(\fnu, \ft) |^2$, the secondary spectrum,
where the tilde  indicates a two dimensional Fourier transform
and $\fnu$ and $\ft$ are variables conjugate 
to $\nu$ and $t$, respectively.
%%bjr added sentence about finite ranges:
The total receiver bandwidth and integration time
define finite ranges for the transform.

\begin{figure}[htb]
\epsscale{0.45}
%\plotone{dynsec_examples.ps}
%\plotone{fig1new.eps}
\plotone{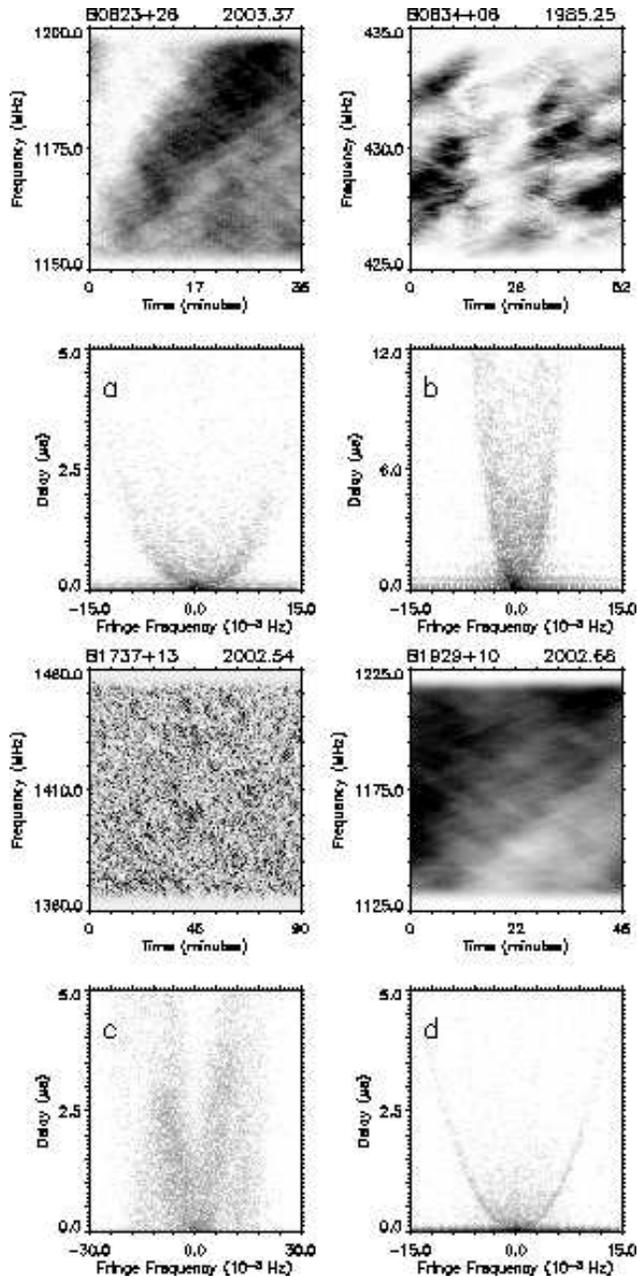}
\figcaption[]{
Primary and secondary spectrum pairs are shown for 4 pulsars that
exhibit the scintillation arc phenomenon.  
%As is true for the other figures in this section, 
The grayscale for the primary
spectrum is linear in flux density.  For the secondary spectrum, the logarithm
of the power is plotted, and the grayscale extends from 3 dB above the noise
floor to 5 dB below the maximum power level.  The dispersion measures 
of these pulsars range from 3.2 to 48.4~pc~\rm cm$^{-3}$ 
and the scattering measures range from 
$10^{-4.5}$ to $10^{-3.6}$ kpc~m$^{-20/3}$, with 
PSR~B1929+10 and PSR~B0823+26 representing the 
extremes in scattering measure.  The data shown here and in 
Figures~\ref{fig:0834_multifreq}-\ref{fig:1133_variety} were obtained from
the Arecibo Observatory.
\label{fig:dynsec_examples} }
\end{figure}

\begin{figure}[htb]
\epsscale{0.95}
%\plotone{0834_multifreq.ps}
%\includegraphics[angle=90,width=16cm]{0834_multifreq.ps}
%\includegraphics[angle=90,width=16cm]{fig2new.eps}
%\plotone{fig2new.eps}
\plotone{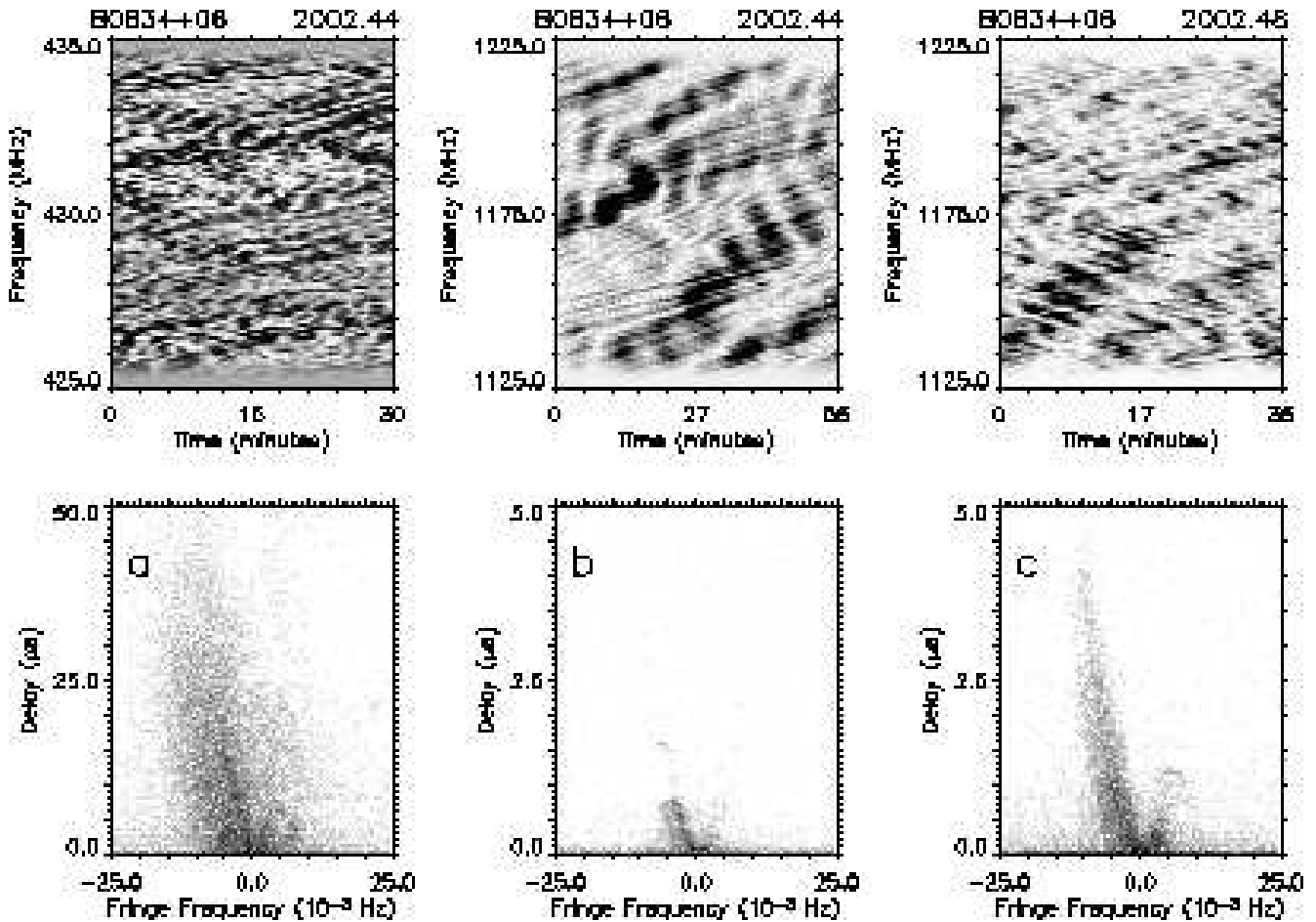}
\figcaption[]{Three observations of PSR~B0834+06 taken within two weeks of each other.  The asymmetry with respect to the conjugate time axis is present, 
in the same sense, in all three observations.  The broad power 
distribution at 430~MHz in (a) is much sharper one day later at 
1175~MHz ($b$);   
however a more diffuse component has returned 14 days later  ($c$).   
Note that the scales for the delay axis in ($b$) and ($c$) differ
from that in ($a$).
The inverted parabolic arclets noticeable in panels $b$ and $c$ are 
a common form of substructure for this and several other pulsars. 
\label{fig:0834_multifreq} }
\end{figure}

\begin{figure}[htb]
\epsscale{0.95}
%\plotone{1133_variety.eps}
%\plotone{1133_variety.rotate.ps}
%\plotone{fig3new.eps}
\plotone{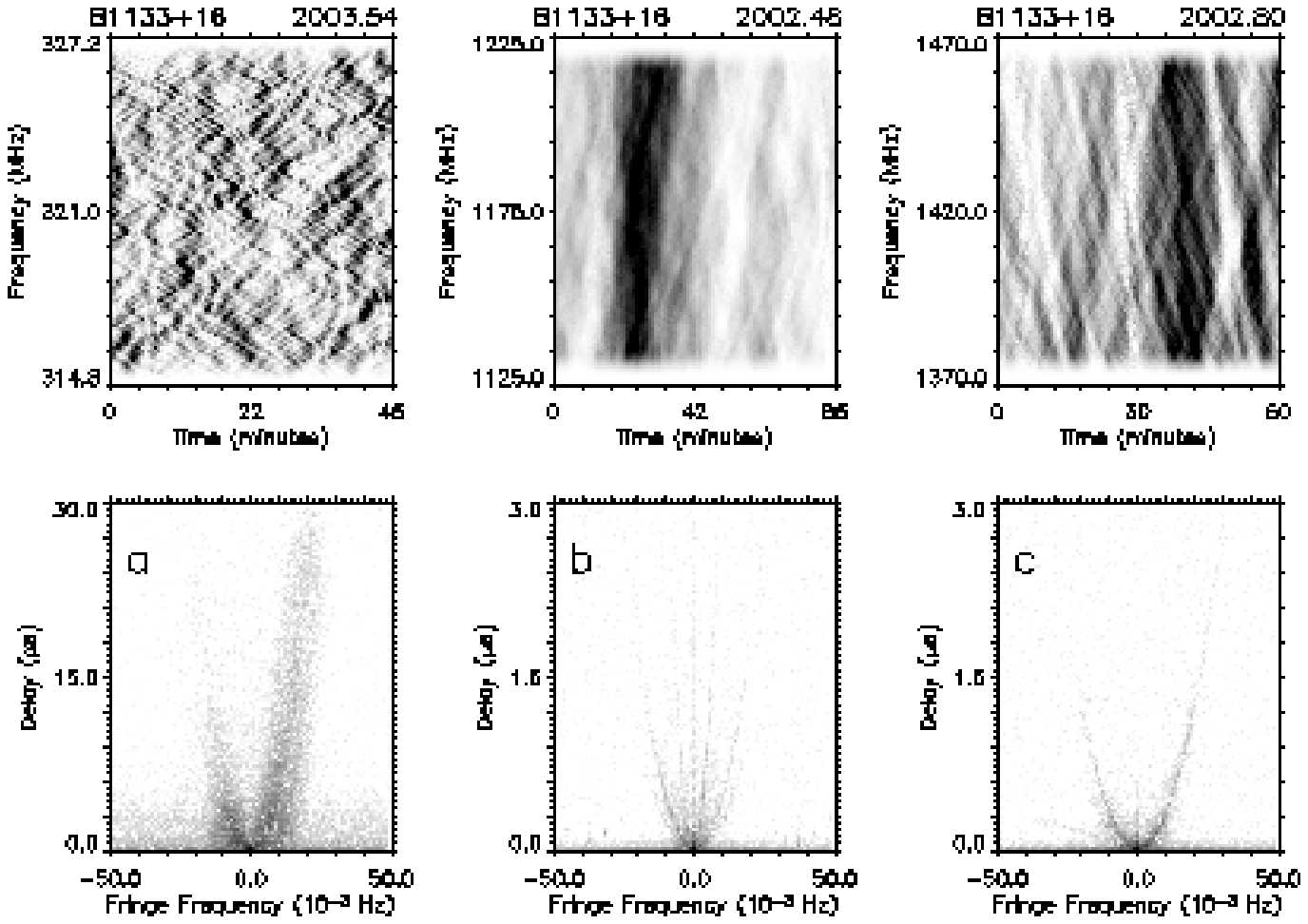}
\figcaption[]{
One of the pulsars in our sample, PSR~B1133+16, shows multiple
scintillation arcs on occasion.  The broad, asymmetric power distribution 
in ($a$) has numerous arclets at 321~MHz. Panels ($b$) and ($c$) are at
frequencies above 1 GHz.   Panel ($b$) shows two clear arcs 
(along with a vertical line at $\ft$ due to narrowband RFI and the sidelobe 
response of power near the origin).  Four months later($c$), only the 
outer of these two arcs -- widened by the $a \propto \nu^{-2}$ scaling
-- is visible.
\label{fig:1133_variety} }
\end{figure}

The dynamic spectrum shows randomly distributed
diffractive maxima (scintles) in the frequency-time plane.
Examples are shown in Figure~\ref{fig:dynsec_examples}.  
In addition, organized patterns such as tilted scintles, 
periodic fringe patterns, and a loosely organized crisscross
pattern have been observed and studied
(e.g. Hewish, Wolszczan, \& Graham 1985; Cordes \& Wolszczan 1986).
In an analysis of four bright pulsars observed 
with high dynamic range at Arecibo, we discovered that
the crisscross pattern  has a distinctive  signature in the 
secondary spectrum (Paper~1).
Faint, but clearly visible power extends 
away from the origin in a parabolic
pattern or arc, the curvature of which depends on observing
frequency and pulsar.  
The observational properties of these 
{\it scintillation arcs} are explored in
more detail in Hill \etal 
(2003; hereafter Paper~2) and Stinebring \etal 
(in preparation, 2004; hereafter Paper~3). 
Here we summarize the major properties
of pulsar observations in order to provide a context for their 
theoretical explanation: 

\begin{enumerate}
\item  Although scintillation arcs are faint, 
they are ubiquitous and persistent when the secondary
spectrum has adequate dynamic range 
\footnote{A signal-to-noise ratio of about 
$10^3$ in the secondary spectrum is usually 
sufficient to show scintillation arcs.   Of 12 bright
($S_{400} > 60$~mJy) pulsars that we have observed at Arecibo, 
10 display scintillation arcs, and 
the other two  have features related to arcs.}
and high frequency and time resolution 
(see Figure~\ref{fig:dynsec_examples} and  Paper~3).
\item The arcs often have a sharp outer 
edge, although in some cases (e.g. Figure~\ref{fig:dynsec_examples}c) 
the parabola is a guide to a diffuse power 
distribution.  
There is usually power inside the parabolic arc 
(Figure~\ref{fig:dynsec_examples}), but the power falls off rapidly outside the arc
except in cases where the overall distribution is diffuse.
\item The  arc outlines are parabolic with
minimal tilt or offset from the origin:  $\fnu = a \ft^2$.   
Although symmetrical outlines are the norm, there 
are several examples of detectable shape asymmetries 
in our data. 
\item In contrast to the symmetrical shape typical of the arc outline, the power distribution of the secondary spectrum can be highly asymmetric
in $\ft$ for a given $\fnu$ and can show significant substructure.
An example  is shown in 
Figure~\ref{fig:0834_multifreq}.
The timescale for change of this substructure is not well established, but some
patterns and asymmetries have persisted for months.
\item A particularly striking form of substructure consists of inverted
{\it arclets} with the same value of  $\vert a \vert$ and with apexes
that  lie along or inside the main arc outline
(Figure~\ref{fig:0834_multifreq} panels b and c). 
\item 
Although a single scintillation arc is usually present for each pulsar, there is one case (PSR~B1133+16) in which multiple 
scintillation arcs, with different $a$ values, are seen 
(Figure~\ref{fig:1133_variety}).  At least two distinct $a$ values 
(and, perhaps, as many as four) are traceable over decades of time. 
\item 
Arc curvature accurately follows a simple scaling 
law with observing frequency:  $a \propto \nu^{-2}$  
(Paper~2).  In contemporaneous 
month-long observations over the range 0.4 -- 2.2~GHz, 
scintillation arcs were present at all frequencies if they
were visible at any for a given pulsar, and the scintillation 
arc structure became sharper and better defined at high frequency.
\item 
The arc curvature parameter $a$ 
is constant at the  5--10\% level for $\sim 20$~years for the 
half dozen pulsars for which we have long-term data spans
(see Paper~3). 
%(e.g. Figure~\ref{fig:1133kappa}).

\end{enumerate}
In this paper we explain or otherwise address all of these points
as well as explore the dependence of arc features on properties
of the scattering medium and source.

\section{Theory of Secondary Spectra and Scintillation Arcs}
\label{sec:theory}

\subsection{A Simple Theory}
\label{sec:basics}

%The basic properties of the parabolic arcs can be described by a
%simple physical model based on the angular spectrum of
%scattered radiation, i.e.\ the apparent brightness distribution
%$B(\thvec)$. We start with a plane wave incident on 
%a thin phase changing screen at distance $z$ from an observer,
%and 

Though the parabolic arc phenomenon is striking and
unexpected, it can be understood in terms of a simple 
model based on a ``thin screen'' of ionized gas containing fluctuations 
in electron density on a wide range of scales.  Incident radiation
is scattered by the screen and then 
%into an angular spectrum of plane waves, 
travels through free space to the observer, at whose location radiation
has a distribution in angle of arrival.  
The intensity fluctuations are caused by interference between 
different components of the angular spectrum whose 
relative phases increase with distance from the screen. 
Remarkably, we find that the arcs can be explained solely 
in terms of phase differences from geometrical path lengths.  The phenomenon
can equally well be described, using the Fresnel-Kirchoff 
diffraction integral, by spherical waves originating 
at a pair of points in the screen interfering at the observer
having traveled along different geometric paths.
Thus, though the screen is dispersive, the arcs
arise simply from diffraction and interference.
Dispersion and refraction in the screen will likely alter the
shapes of the arcs, though we do not analyze these effects in this
paper.

To get interference, the screen must be illuminated by a radiation field
of high spatial coherence, i.e. radiation from a point-like source.
The source can be temporally incoherent because the different 
components of the temporal Fourier transform each contribute 
nearly identical interference patterns, as in interferometry. 
Two components of the
angular spectrum arriving from directions $\thvec_1$ and $\thvec_2$
interfere to produce a two-dimensional fringe pattern 
%bjr1
whose phase varies slowly with observing frequency. The pattern is sampled
in time and frequency as the observer moves through it,
creating a sinusoidal fringe pattern in the dynamic spectrum. 
The fringe appears as a single Fourier component in the secondary spectrum.
Under the small-angle scattering of the ISM its $\fnu$ coordinate is 
the differential geometric delay 
$ \propto \theta_2^2 - \theta_1^2$ and its $\ft$ coordinate
is the fringe rate $\propto \vperpvec \cdot (\thvec_2 - \thvec_1)$,
where $\vperpvec$ 
is an appropriate transverse velocity (see below).
A quadratic relationship $\fnu$ to $\ft$ results 
naturally from their quadratic and linear dependences on the angles.
When one of the interfering waves is undeviated (e.g.
$\thvec_1 = 0$) we immediately get the simple parabola
$\fnu \propto \ft^2$. 
%%bjr
When the scattering is weak there will be a significant 
undeviated contribution resulting in a simple parabolic arc.
However, we also find that an arc appears 
in strong scattering due to the interference of waves
contained within the normal scattering disc
with a faint halo of waves scattered at much larger
angles.  The faint halo exists only for scattering media having 
 wavenumber spectra less steep than (wavenumber)$^{-4}$, including 
the Kolmogorov spectrum.

While the relation of $\fnu$ to time delay is well known 
through the Fourier relationship of the pulse broadening 
to the radio spectrum intensity scintillation, the physical
interpretation of $\ft$ can be viewed in several ways besides the
fringe rate. Scintillation versus time is often Fourier analyzed into 
a spectrum versus frequency ($\ft$) in Hz; in turn this is simply related to
spatial wavenumber $\ft=\kappavec \cdot \vperpvec /(2\pi)$ 
and hence to angle of arrival ($\kappavec = k \thvec$). 
It can also be thought of as the beat frequency due to the
different Doppler shifts of the two scattered waves.  
Thus the secondary spectrum can be considered as a 
differential \it delay-Doppler spectrum \rm, similar to that measured in radar
applications.
%bjr2
%bjr  This sentence seems to belong further on:
% For arcs to be manifested, however, 
% particular distributions of the angular spectrum of scattered radiation are
% required.  

\subsection{Screen Geometry}
\label{sec:screen-theory}

In this section we discuss the relationship between the scattered image
and the secondary spectrum $S_2$.
Later, in \S\ref{sec:scattphys},
we obtain the relation more rigorously by   
deriving $S_2$ as a fourth moment of the electric
field.  Explicit results can be obtained in the limits 
of strong and weak scintillation,
as shown in Appendices \ref{sec:app-strong} and 
\ref{sec:app-weak}. 
We find that the weak scintillation result is simpler,
being a second moment of the electric field
since it only involves the interference of 
scattered waves with the unscattered field.
However, pulsars are typically observed in strong scintillation,
and the strong scintillation limit gives exactly the same
result as in the approximate theory (Eq.~\ref{eq:fpq1})
used below. So we now apply the approximate theory 
to spherical waves from a pulsar scattered
by a thin screen.

Consider the following geometry: a point source at $z=0$,
a thin screen at $z=\ds$ and an observer at $z=D$.
For convenience, we define $s = \ds /D$.
%where $\rsvec, \rvecp$ and $\rvec$ are two dimensional vectors.
The screen changes only the phase of incident
waves, but it can both diffract and refract radiation from the source. 
%bjr1
For a single frequency emitted from the source 
%bjr2
two components of the angular spectrum at
angles $\thvec_1$, $\thvec_2$ (measured relative to the direct path)
sum and interfere at the observer's location with a 
phase difference $\phiij$. 
The resulting intensity 
is  $ I = I_1 +  I_2 + 2 \sqrt{I_1 I_2}\cos \phiij$
where $I_1$ and $I_2$ are the intensities from 
each component  (e.g. Cordes \& Wolszczan 1986). 
The total phase difference 
$\phiij = \Phi_{\rm g} + \phi$,
includes a contribution from 
geometrical path-length
differences $\Phi_{\rm g}$ and from the screen phase, $\phi$, 
which can include both small and large-scale structures that refract
and diffract radiation.
Expanding $\phiij$ to first order in time and frequency
increments, $\delta t$ and $\delta\nu$, the phase difference is
\be
\phiij = \Phi_0 + 2\pi (f_t \delta t + f_{\nu} \delta\nu),
\ee
where $\delta\nu = \nu - \nu_0$, $\delta t = t - t_0$ and 
$t_0,\nu_0$ define the center of the observing window; 
$f_t = (1/2\pi) \partial_t \phiij$ is the  
fringe rate or differential Doppler shift, and 
$f_{\nu}=(1/2\pi) \partial_{\nu} \phiij$ is the differential group delay.
In general $\Phi$ includes both a geometrical path-length difference 
and a dispersive term. At the end of \S\ref{sec:ISconditions}
we consider the effects of dispersion, but in many cases 
of pulsar scintillation it can be shown that
the dispersive term can be ignored.  
We proceed to retain only the geometric delays and obtain results 
(Appendix \ref{sec:appendixa}) that were reported in Paper 1:
\be
f_{\nu} &=&  \left[ \frac{D(1-s)}{2\,c\,s }\right]
		\left(\thvec_2^2 - \thvec_1^2\right) 
	\label{eq:fnu} \\
f_t &=& 
  \left(\frac{1}{\lambda\, s}\right)(\thvec_2 - \thvec_1)\cdot\vperpvec. 
	\label{eq:ft}	
\ee
%bjr1
Here $\lambda$ is the wavelength at the center of
the band, $s=\ds/D$, and $\vperpvec$ is the velocity of the point in the screen
intersected by a straight line from the pulsar to the observer,
given by a weighted sum of the velocities
%bjr2
of the source, screen and observer (e.g. Cordes \& Rickett 1998):
\be
\vperpvec = (1-s)\vpperpvec + s \vobsperpvec  
-\vismperpvec  .   
\label{eq:veff}
\ee
It is also convenient to define an effective
distance $\dse$, 
\be
\dse=Ds(1-s) \;.
%\dse=\ds(1-s) \;.
\label{eq:defdse}
\ee

The two-dimensional Fourier transform of 
the interference term $\cos\phiij(\delta\nu,\delta t)$
is a pair of delta functions
placed symmetrically about the origin of the 
delay fringe-rate plane. The secondary spectrum 
% bjr1
at ($\fnu,\ft $) is thus the summation of the delta functions 
from all pairs of angles subject to Eq.~\ref{eq:fnu} 
\& \ref{eq:ft}, as we describe in the next section.

%
%Use of delta functions is an idealization, of course, because
%real data are obtained over a finite time and bandwidth.   The
%features in the secondary spectrum therefore have finite width and
%amplitude,  which we discuss further in \S\ref{sec:app-num}.  
%\subsection{Derivation of the Secondary Spectrum} %bjr  
% I changed the name since this section only mentions 
% methods of derivation but then uses
% the brightness integral without deriving it formally.

\subsection{Secondary Spectrum in Terms of the Scattered Brightness} 
\label{sec:spec2}

%bjr1   substantial changes all through here....
%
%\bf bjr  removed the equation that was already introduced in 
%the Observation section. Also made multiple changes see .tex file for 
%more details \rm
% I am removing it since it is not used in what follows
% if we want to get formal we should but in the finite range of
% frequency and time,.... which is done later.
%
%Operationally, a measured dynamic spectrum $S(\nu, t)$ 
%has a secondary spectrum
%\be
%S_2(\fnu, \ft) = \vert \tilde S(\fnu, \ft) \vert^2,  
%\ee
%where $\tilde S(\fnu, \ft)$ is the two-dimensional Fourier transform
%of the dynamic spectrum  and the angular brackets denote an 
%average.   

In this section we examine how the form of the secondary spectrum
varies with the form assumed for the scattered image,
without considering the associated physical conditions in the medium.
We postpone to \S\ref{sec:scattphys} a discussion of the
scattering physics and the influence of the integration
time.  The integration time is important in determining
whether the scattered brightness is a smooth function or
is broken into speckles as discussed by Narayan \& Goodman (1989).
This in turn influences whether the secondary spectrum
takes a simple parabolic form or becomes fragmented.

We analyze an arbitrary scattered image by treating its scattered 
brightness distribution, $B(\thvec)$,  as a probability
density function (PDF) for the angles of scattering. 
In the continuous limit the secondary spectrum is the joint 
PDF of $\fnu,\ft$ subject to the constraints of
Eq.~(\ref{eq:fnu} \& \ref{eq:ft}). 
It is convenient to use dimensionless variables for the
delay and fringe rate:
\be
p &=& \thvec_2^2 - \thvec_1^2
        =  \left [\frac{2c s}{D(1-s)}\right ] \fnu,
\label{eq:pbar}\\
q &=& (\thvec_2 - \thvec_1)\cdot\vperpvechat
        = \left(\frac{\lambda s}{\vperp}\right) \ft,
\label{eq:qbar}
\ee
where $\vperpvechat$ is a two-dimensional unit vector for the transverse
effective velocity.    
It is also useful to normalize angles by the characteristic
diffraction angle, $\thd$, so in some contexts discussed below 
$\theta\to\theta / \thd$, in which case $p\to \fnu /\taud$ and 
$q\to 2\pi \dtd \ft$, where $\taud$ is the
% $C_1 \sim 1$ is defined by the relationship between the 
pulse broadening time and $\dtd$ is the diffractive 
scintillation time.
% and diffraction bandwidth $\dnud$:
% $2\pi\taud\dnud = C_1$ (Cordes \& Rickett 1998). 
The secondary spectrum is then given by an integral
of the conditional probability $\delta(p-\pbar)\delta(q-\qbar)$
for a given pair of angles multiplied by the PDFs of those angles,
\be
\fpq(p,q) = \int\int d\thvec_1\,d\thvec_2\,
        \fthvec(\thvec_1) \fthvec(\thvec_2)
        \delta(p-\pbar)\delta(q-\qbar),
\label{eq:fpq1}
\ee
where $\pbar,\qbar$ are the values of $p,q$ for particular values
of $\thvec_{1,2}$ as given in Eq.~\ref{eq:pbar} and \ref{eq:qbar}.
In Appendix \ref{sec:app-strong}, Eq.~\ref{eq:fpq1} is 
derived formally from the ensemble average in the limit of strong
diffractive scattering.

The secondary spectrum is symmetric through the origin
($p\to -p$ and $q\to -q$).  Eq.~(\ref{eq:fpq1}) shows
that it is essentially 
a distorted autocorrelation 
%bjr1 operation,
%containing features that in some cases are 
%similar to those in the autocorrelation 
of the scattered image.  
%bjr2
With no loss of generality we simplify the analysis 
by taking the direction of the 
velocity to be the $x$ direction.
%i.e. $\vperpvechat = \hat {\xvec}$.
The four-fold integration in Eq.~\ref{eq:fpq1} may be 
reduced to a double integral by integrating the delta functions
over, say,  $\thvec_2$ 
%and let $\thvec_1 = (x,y)$ 
to obtain
\be
\fpq(p, q) = \half\int\int d{\theta_1}_x\, d{\theta_1}_y\,
	H(U) U^{-1/2} \fthvec({\theta_1}_x, {\theta_1}_y) 
	\left [ \fthvec(q + {\theta_1}_x, \sqrt{U}) + 
     \fthvec(q+{\theta_1}_x, -\sqrt{U}) 
	\right ],
\label{eq:fpqU}
\ee
where 
\be
U \equiv p - q^2 - 2q{\theta_1}_x + {\theta_1}_y^2  
\label{eq:udef}
\ee
and $H(U)$ is the unit step function.   
With this form,  the integrand is seen to maximize at the singularity
$U = 0$, which yields a quadratic relationship between $p$ and $q$.   
For an image offset from the origin by angle $\thvec_0$,
e.g. $\fthvec(\thvec)\to \fthvec(\thvec-\thvec_0)$, the form
for $\fpq$ is similar except that 
$U = p - q^2 -2q(\theta_{1x} +{\theta}_{0x}) + 
({\theta_1}_y + {\theta}_{0y})^2$ and
the ${\theta}_y$ arguments of $B$ in square brackets in Eq.~\ref{eq:fpqU} 
become $\pm\sqrt{U}-{\theta}_{0y}$.  
The integrable singularity at $U=0$ makes the form of  Eq.~\ref{eq:fpqU}
inconvenient for numerical evaluation. In Appendix B we 
use a change of variables to avoid the singularity,
giving form (Eq.~\ref{eq:fpqxy}) that 
can be used in numerical integration.  However, this does
not remove the divergence of $\fpq$ at the origin $p=q=0$.
In Appendix \ref{sec:app-num} we show that inclusion of 
finite resolutions in $p$ or $q$, which follow from the finite
extent of the orignal dynamic spectrum in time and frequency, 
avoids the divergence at the origin,
emphasizing that the dynamic range
in the secondary spectrum is strongly influenced by
resolution effects.

%\subsubsection{Scaling Law for Parabolic Arcs}

The relation $p=q^2$ follows from the singularity $U=0$ after
integration over angles in Eq.~\ref{eq:fpqU} and is confirmed in  
a number of specific geometries discussed below.  This relation becomes
$\fnu = a \ft^2$ in dimensional units, with
\be
a = \frac{Ds(1-s)}{2c}\left(\frac{\lambda}{\vperp} \right)^2
  = 0.116
\left [ \frac{s(1-s)}{1/4} \right] 
D_{\rm kpc}\nu^{-2} V_{100}^2
\,{\rm sec^{-3}},\,
\label{eq:a}
\ee
where $\vperp = 100\,{\rm km\,s^{-1}}\, V_{100}$ and $\nu$ is in GHz.
For screens halfway between source and observer ($s=1/2)$,
this relation  yields values for $a$ that are 
consistent with  the arcs
evident in Figures~\ref{fig:dynsec_examples}-\ref{fig:1133_variety}
and also shown in Papers 1-3.  As noted in Paper 1, $a$ 
does not depend on any aspect of the scattering screen save for
its fractional distance from the source, $s$, and it 
maximizes at $s=1/2$ for fixed $\vperp$.  
When a single arc occurs and other parameters
in Eq.~\ref{eq:a} are known, $s$ can be determined to within a pair
of solutions that is symmetric about $s=1/2$. However, when $\vperp$
is dominated by the pulsar speed, $s$ can be determined uniquely
(c.f. Paper 1).

\subsection{Properties of the Secondary Spectrum}
\label{sec:properties}
To determine salient properties of the secondary spectrum that
can account for many of the observed phenomena, we consider special 
cases of images for which Eq.~\ref{eq:fpq1} can be evaluated.
As noted above the effects of finite resolution are important both
observationally and computationally and are discussed in 
Appendix \ref{sec:app-num}.  

%\paragraph*{Point Images}
\subsubsection{Point Images}

A point image produces no interference effects, so the 
secondary spectrum consists of a delta function at the origin,
$(p,q) = (0,0)$.    
Two point images with amplitudes $a_1$ and $a_2$
produce fringes in the dynamic spectrum, which give
delta functions of amplitude  $a_1a_2$ at position given by 
Eq.~\ref{eq:pbar} and \ref{eq:qbar} and its counterpart 
reflected through the origin and also a 
delta function at the origin with amplitude $1-2a_1a_2$. 
Evidently, an assembly of point images gives a
symmetrical pair of delta functions in $p, q$
for each pair of images.

%\paragraph*{One-dimensional Images}
\subsubsection{One-dimensional Images}

Arc features can be prominent for images elongated
in the direction of the effective velocity. Consider an 
extreme but simple case of a one dimensional image extended along
the $x$ axis only
\be
\fthvec(\thvec) =  g(\theta_x)\delta(\theta_y), 
\label{eq:1dx}
\ee
and where $g$ is an arbitrary function. The secondary spectrum is
\be
\fpq(p, q) = \left (2\qabs\right)^{-1}
	g \left( \frac{p-q^2}{2q} \right ) 
	g \left( \frac{p+q^2}{2q} \right ). 
\label{eq:fpq1D}
\ee
By inspection, parabolic arcs extend along
$p = \pm q^2$ with amplitude 
% jmc: $\fpq \propto g(|q|)/|q|$: arg of g should be q not |q|.
$\fpq \propto g(q)/|q|$
along the arc, becoming 
% narrower 	jmc 20 June 2004 changed narrower -> wider. 
wider in $p$ at large $|q|$.  However, for the 
particular case where $g$ is a Gaussian function, the product
of the $g$ functions in Eq.~~(\ref{eq:fpq1D}) 
$\propto \exp[-(p^2 + q^4)/2q^2]$, which cuts off
the arcs very steeply (see \S\ref{sec:gauss}).  
For images with slower fall offs at large angles,
the arc features can extend far from the origin.
For the same image shape elongated transverse rather than
parallel to $x$, inspection of Eq.~\ref{eq:qbar} - \ref{eq:fpq1} 
indicates that
$\fpq(p, q) \propto \delta(q)$, so there is only a
ridge along the $q$-axis. 
These examples suggest that prominent arcs are expected
when images are aligned with the direction of the effective velocity,
as we confirm below for more general image shapes.

%\paragraph*{Images with a Point and an Extended Component} 
\subsubsection{Images with a Point and an Extended Component} 
\label{sec:pt+ext}

Consider a scattered image consisting of a point source at
$\thvec_p$ and an arbitrary, two-dimensional image component, $g(\thvec)$,
\be
\fthvec(\thvec) = a_1 \delta(\thvec - \thvec_p) + a_2 g(\thvec).
\label{eq:pplusg}
\ee
The secondary spectrum consists of the two self-interference terms from
each image component and cross-component terms of the form
\be
\Delta \fpq(p,q) = 
	\half a_1 a_2 U^{-1/2} H(U)
	%\left [
	\sum_{\pm}
	g(q+{\theta_p}_x, \pm\sqrt{U})
	%g(q+{\theta_p}_x, \sqrt{U}) + g(q+{\theta_p}_x, -\sqrt{U}) 
	+ S.O.,
	%\right ],
\label{eq:fpqg1}
\ee
where $U = p - q^2 - 2q{\theta_p}_x + {\theta_p}_y^2$, 
$H(U)$ is the unit step function and
``S.O.'' implies an additional term that is symmetric through
the origin, corresponding to 
letting $p\to - p$ and $q\to -q$ in the first component. 
A parabolic arc is defined by the $H(U) U^{-1/2}$ factor. 
As $U \to 0$ the arc amplitude 
is $a_1 a_2 U^{-1/2} g(q+{\theta}_{p_x}, 0)$. Considering
$g$ to be centered on the origin with width
${W_{\rm g}}_x$ in the $x$ direction, the arc extends to 
$\vert q + {\theta_p}_x \vert \lesssim {W_{\rm g}}_x/2$.
Of course if the point component has a small
but non-zero diameter, the amplitude of the arc
would be large but finite. If the point is at
the origin, the arc is simply $p=q^2$ as already discussed.
In such cases Eq.~\ref{eq:fpqg1} can be inverted to estimate 
$g(\thvec)$ from measurements of $\Delta\fpq$.
The possibility of estimating the two-dimensional 
scattered brightness from observations with a single dish
is one of the intriguing aspects of the arc phenomenon.

With the point component displaced from the origin, the apex
of the parabola is shifted from the origin to 
\be
(p, q)_{\rm apex} = (-\theta_p^2, -{\theta_p}_x).
\label{eq:apex}
\ee
By inspection of Eq.~\ref{eq:fpqg1}, 
the condition $U=0$ implies that the arc is 
dominated by contributions from image components 
scattered parallel to the velocity vector at angles
$\thvec = (q+{\theta}_{p_x}, 0)$.
%% bjr I removed the italicized sentence it seemed
% to be an un-needed repetition and emphasis
%More generally, arcs emphasize image components 
%scattered at angular displacements
%along a line in the plane of the sky
%that passes through the pulsar's position and is 
%parallel to the transverse velocity vector.
Thus images elongated along the velocity vector produce arcs enhanced over
those produced by symmetric images.
This conclusion is general and is independent
of the location, $\thvec_p$, of the point 
component.

\begin{figure}[htb]
\epsscale{0.7}
\begin{center}
%\plottwo{parabs5-0.ps}{parabs5-3.ps}
%\plottwo{parab0.ps}{parab3.ps}
\plottwo{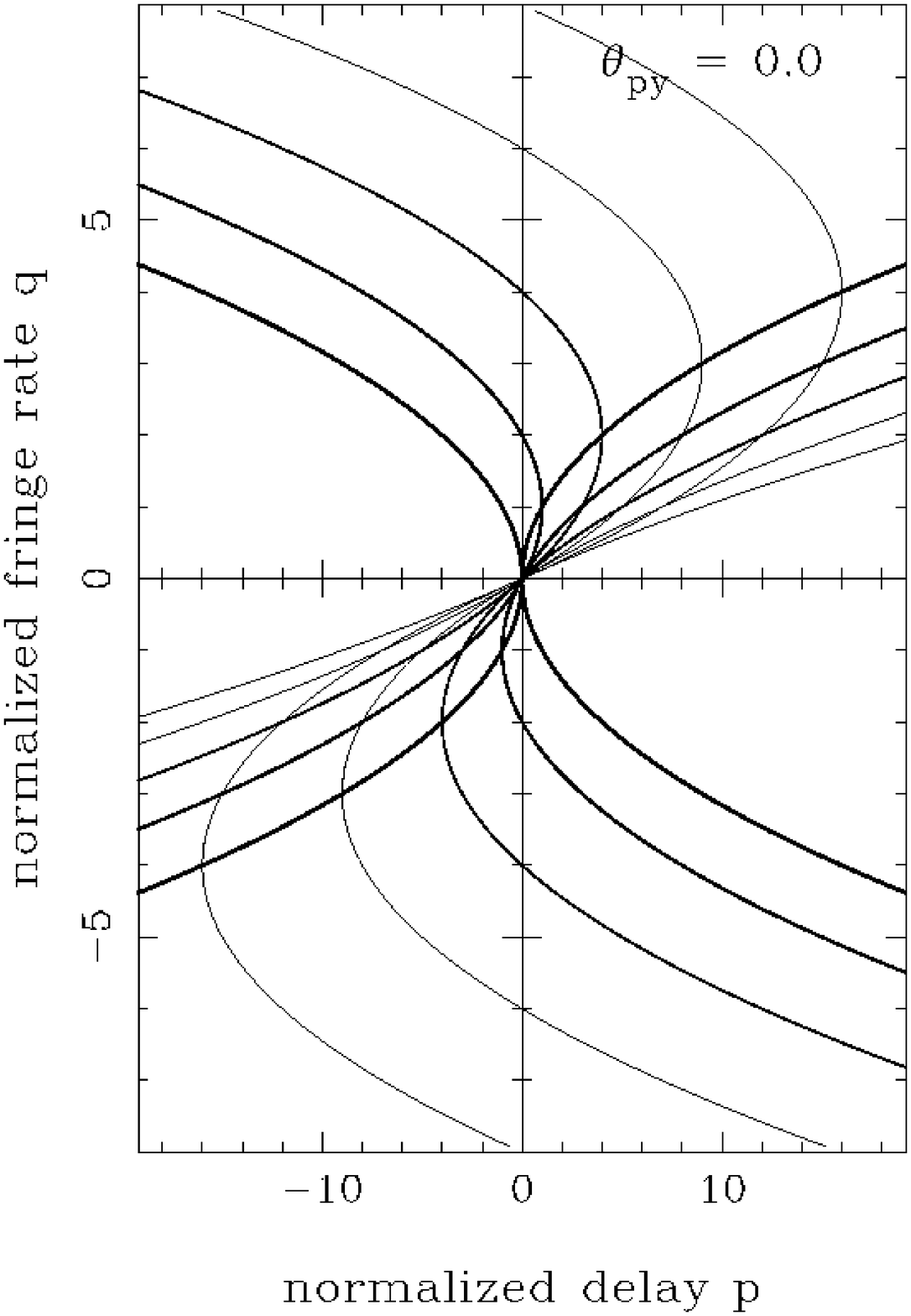}{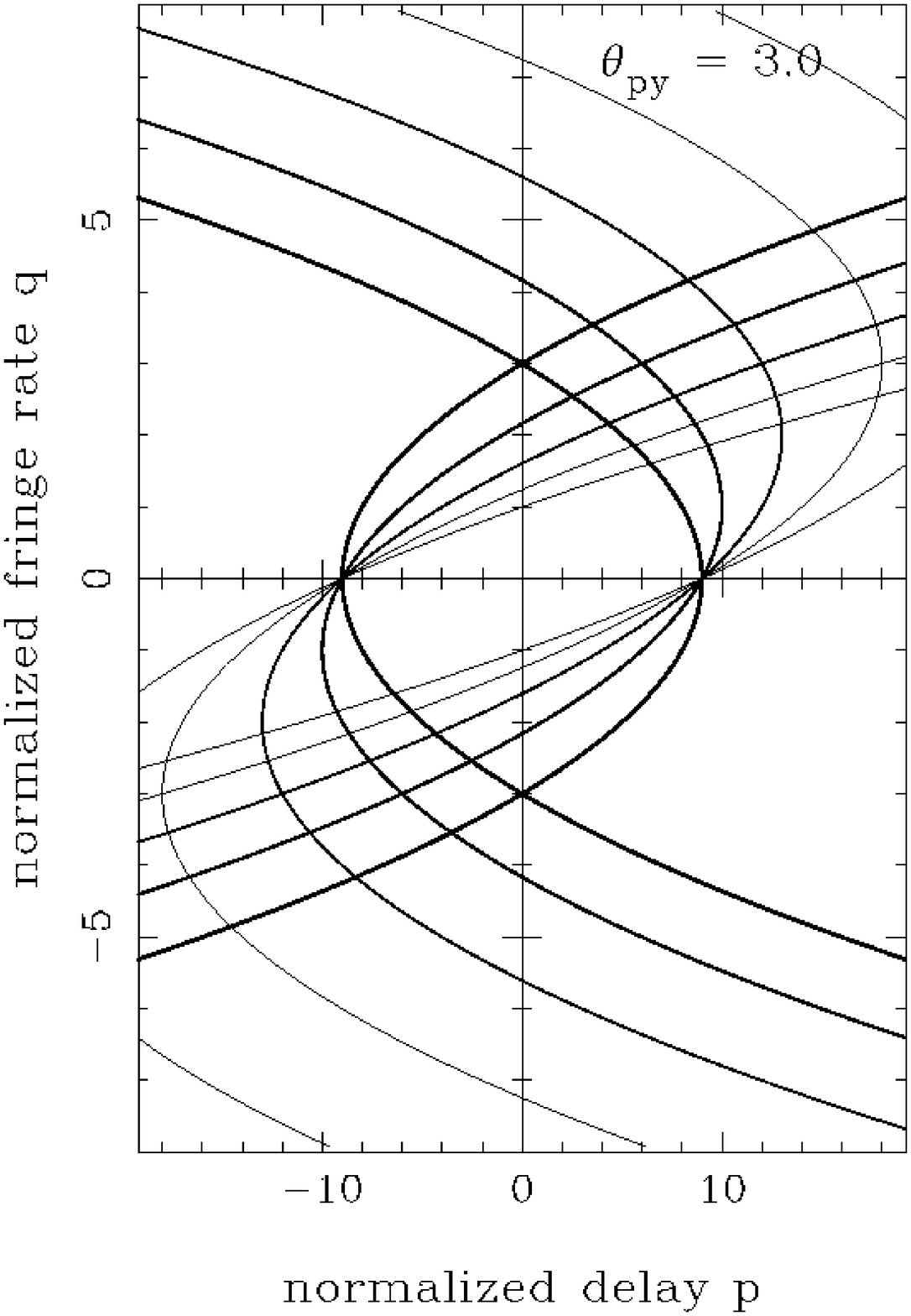}
\end{center}
\figcaption[]{
Trajectories for parabolic arcs due to interference between
a point component  at $(\theta_{px} , \theta_{py})$
with an extended distribution.  The curves are for
five values of normalized  $\theta_{px}$ (0 for thickest line to
4 for thinnest line).  Left panel: for $\theta_{py} = 0$ the parabolas
have apexes of the reverse arcs lying along $p=q^2$. Right panel:
for $\theta_{py} = 3$, the apexes are inside $p=q^2$.
\label{fig:parabs} }
\end{figure}

Figure \ref{fig:parabs} shows parabolic arc lines ($U=0$)
for various assumed positions for the point component.
Eq.\ref{eq:apex} gives a negative delay
at the apex, but since $\fpq$ is symmetric about the origin, 
there is also a parabola with a positive apex, but 
with reversed curvature.
When ${\theta}_{p_y}=0$ the apex must lie on the 
basic arc $p=q^2$ (left panel), 
and otherwise the apex must lie inside it (right panel).
Such features suggest an explanation
for the reversed arclets that are occasionally observed, as
in Figure \ref{fig:0834_multifreq}.
We emphasize that the discussion here has been restricted to 
the interference between a point and an extended
component, and one must add their self-interference
terms for a full description.

%Figure \ref{fig:parabs} shows two sets of parabolae
%for a range of $\psi_{ax}$; in each case the 
%curves all pass through a common point at 
%$p=\pm\psi_{ay}^2$. The part of the parabola that is 
%``illuminated'' is governed by the width and location of the 
%extended component $b_2$. In addition to the
%interference of a point and extended component described
%by  equation (\ref{eq:fpq.interf}) one must add their self-interference
%terms for a full description.

The foregoing analysis can, evidently, be applied to a
point image and an ensemble of subimages.  The general nature
of $\fpq$ is similar in that it is enhanced along the 
curve where $U=0$ which enhances subimages lying
along the velocity vector.

\section{Secondary Spectra for Cases Relevant to the Interstellar Medium}
\label{sec:ismcases}

We now compute the theoretical secondary spectrum for
various simple scattered brightness functions commonly invoked
for interstellar scattering.  The computations include 
self-interference and also the 
finite resolution effects, discussed in Appendix \ref{sec:app-num}.

\subsection{Elliptical Gaussian Images}
\label{sec:gauss}

Measurements of angular broadening of highly scattered OH masers
(Frail et al. 1994), Cyg X-3 (Wilkinson et al. 1994),
pulsars (Gwinn et al. 1993), and
AGNs (e.g. Spangler \& Cordes 1988; Desai \& Fey 2001) 
indicate that the scattering is
anisotropic and, in the case of Cyg X-3, consistent with a Gaussian
scattering image, which probably indicates
diffractive scales smaller than the ``inner scale''
of the medium.  Dynamic spectra cannot be measured for such sources
because the time and frequency scales of the diffractive scintillations are
too small and are quenched by the finite source size.  
However, some less scattered pulsars may have measurable
scintillations that are related to an underlying Gaussian image.

\begin{figure}[h]
\epsscale{0.4}
%\plotone{fpqlist_gau.ps}
\plotone{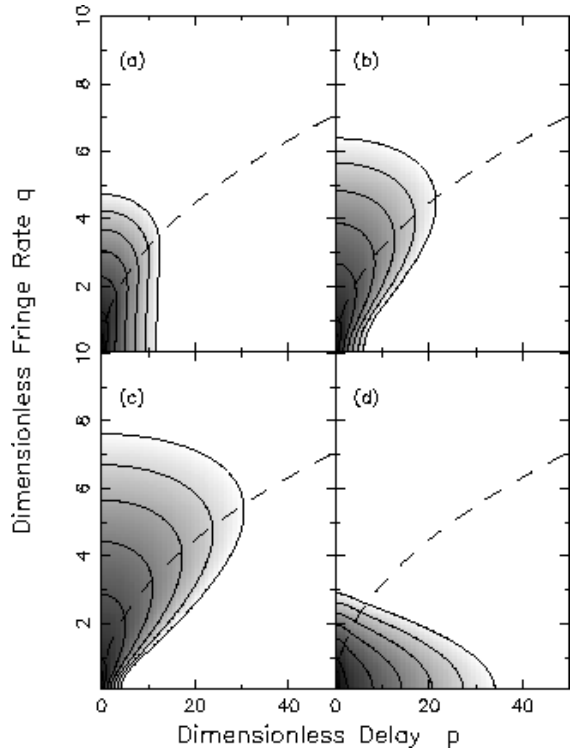}
\figcaption[]{
Secondary spectra corresponding to Gaussian scattered images
with different ellipticities and orientations:
(a) a circular Gaussian image;
(b) an elliptical Gaussian with axial ratio of 2 elongated along 
the $x$ axis;
(c) an axial ratio of 3 elongated along the $x$ axis;
and 
(d) an axial ratio of 3 elongated along the $y$ axis. 
The contour separation is one decade and the outermost contour is
6 decades from the peak of the secondary spectrum.
The dashed line shows the parabola, $p = q^2$.
\label{fig:gaufpqset} }
\end{figure}

The secondary spectrum for an elliptical Gaussian image
cannot be solved analytically, although Eq.~\ref{eq:fpqxy} 
can be reduced to a one dimensional integral 
which simplifies the numerical evaluation.  
Figure \ref{fig:gaufpqset} shows some examples.  The upper left hand panel is
for a single circular Gaussian image. It does 
not exhibit arc-like features, although it ``bulges'' out along 
the dashed arc-line ($p=q^2$).  
Upper right is for an elliptical Gaussian image with a 2:1 axial ratio
parallel to the velocity vector. The 3:1 case (lower left)
shows the deep ``valley'' along the delay ($p$) 
axis, which is characteristic of images elongated parallel 
to the velocity.  Such deep valleys are frequently seen
in the observations and provide evidence for anisotropic scattering.
Notice, however, that the secondary spectrum  can be strong outside of the 
arc line, where the contours become parallel
to the $p$-axis. Inside the arc-line the contours follow curves
like the arc-line.
The 3:1 case with the major axis transverse to the velocity (lower right)
shows enhancement along the delay axis with 
no bulging along the arc-line.

% bjr1 removed this paragraph which refers to diagrams no longer in the paper
%
%{\bf Move next paragraph to later?}
%Lower left is for a circular gaussian image of unit radius,
%which is centered one unit from the origin in $\psi_x$. The result 
%is asymmetry in the fringe-rate corrdinate ($q$), 
%since $q$ responds to differences in
%$\psi_x$. Such asymmetries are seen in the observations,
%but we note that this calculation does not include any frequency
%dependence in the position shift. 
%Lower right has two Gaussian components; one is
%the centered elliptical image as in upper right panel plus a circular
%component displaced by 4 units in $\psi_x$, with a radius
%0.2 and flux density 25\% of the centered component.
%One can see the mutual interference as the reverse arc
%with its apex at $p=16, q=4$, as explained in the next section.
%The thickness of the arc is largely governed by the
%dimensions in $\psi_y$.
% bjr2
 
\subsection{Media with Power-Law Structure Functions}
\label{sec:powerlaw-sfn}

Here we consider brightness distributions associated with 
wavenumber spectra having a power-law form.    
For circularly symmetric scattering, the phase structure
function depends only on the magnitude of the baseline  so the 
visibility function is $\Gamma(b) = \exp[-\half\sfphi(b/s_0)]$.  
A suitable power-law form for the structure function is
$\sfphi(b) =  (b/s_0)^{\alpha}$, 
where $s_0$ is the spatial scale of the intensity 
diffraction pattern, defined such that  
$\vert \Gamma(s_0) \vert^2 = e^{-1}$
(e.g. Cordes \& Rickett 1998). The corresponding image is
\be
%bjr added b db
B(\theta) = 2\pi \int_{0}^{\infty} b \, db \, 
\Gamma(b) J_0(2\pi b \theta / \lambda),
\ee
where $J_0$ is the Bessel function.
Computations are done in terms of an image normalized so that $B(0)=1$,
a scaled baseline $\eta = b / s_0$, and a scaled angular coordinate 
$\psi=\theta / \thd$, where $\thd=(k s_0)^{-1}$ is the 
scattered angular width.
%
%we have
%\be
%I(\psi) = 
%	\frac{
%	\int d\eta  \,\eta \, \Gamma(\eta) J_0(\eta\psi).
%	}
%	{
%	\int d\eta  \,\eta \, \Gamma(\eta) 
%	}
%%I(\theta / \thd) = 
%%\frac{
%%\int d\beta  \,\beta \, \Gamma(\eta) J_0(\beta\theta / \thd).
%%}
%%{
%%\int d\beta  \,\beta \, \Gamma(\beta) 
%%}
%\ee 

%\paragraph*{Single-slope Power Laws }
\subsubsection{Single-slope Power Laws }
\begin{figure}[htb]
\begin{center}
\epsscale{0.7}
%\plottwo{powimageset.ps}{kolimageset.ps}
\plottwo{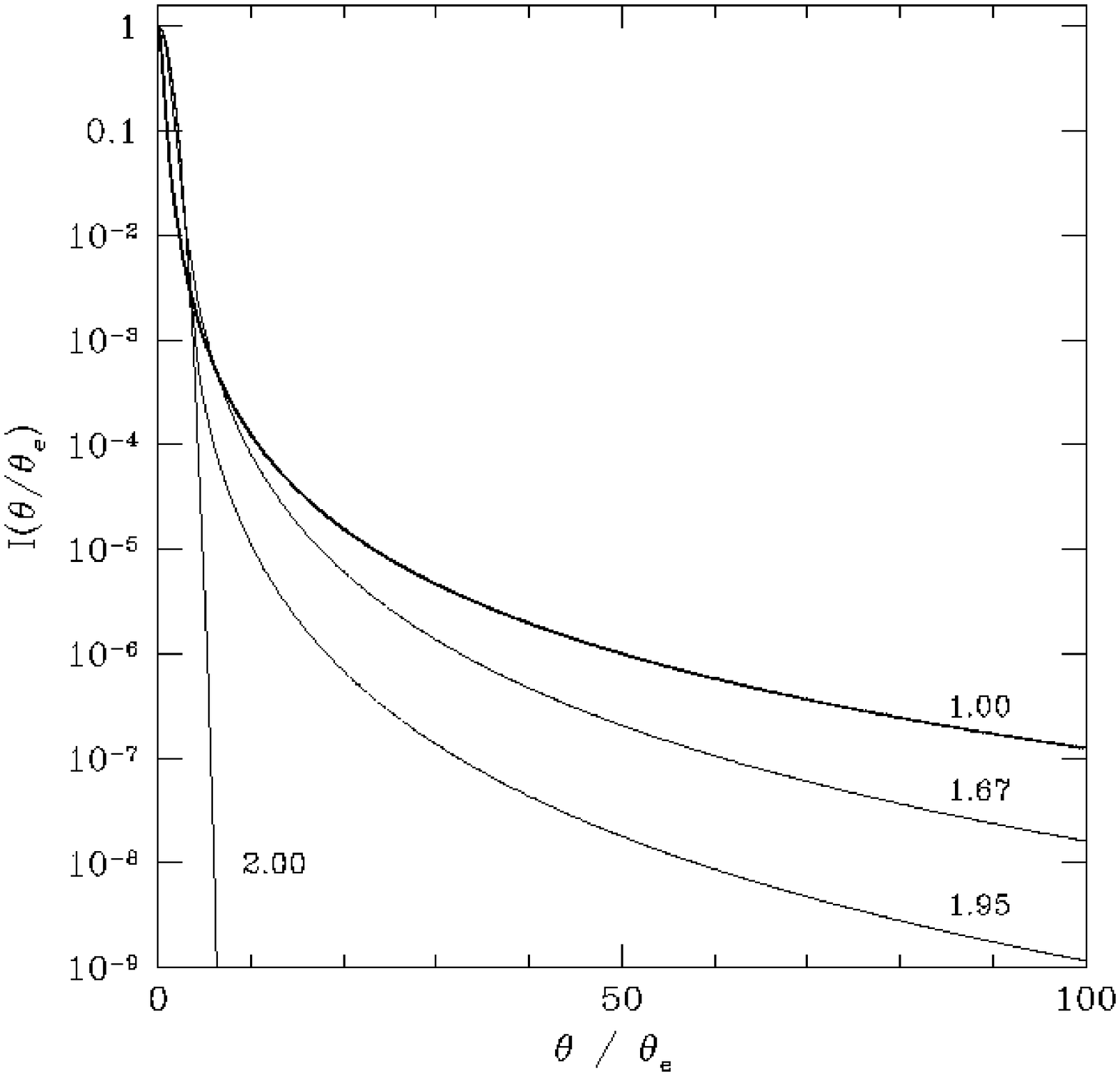}{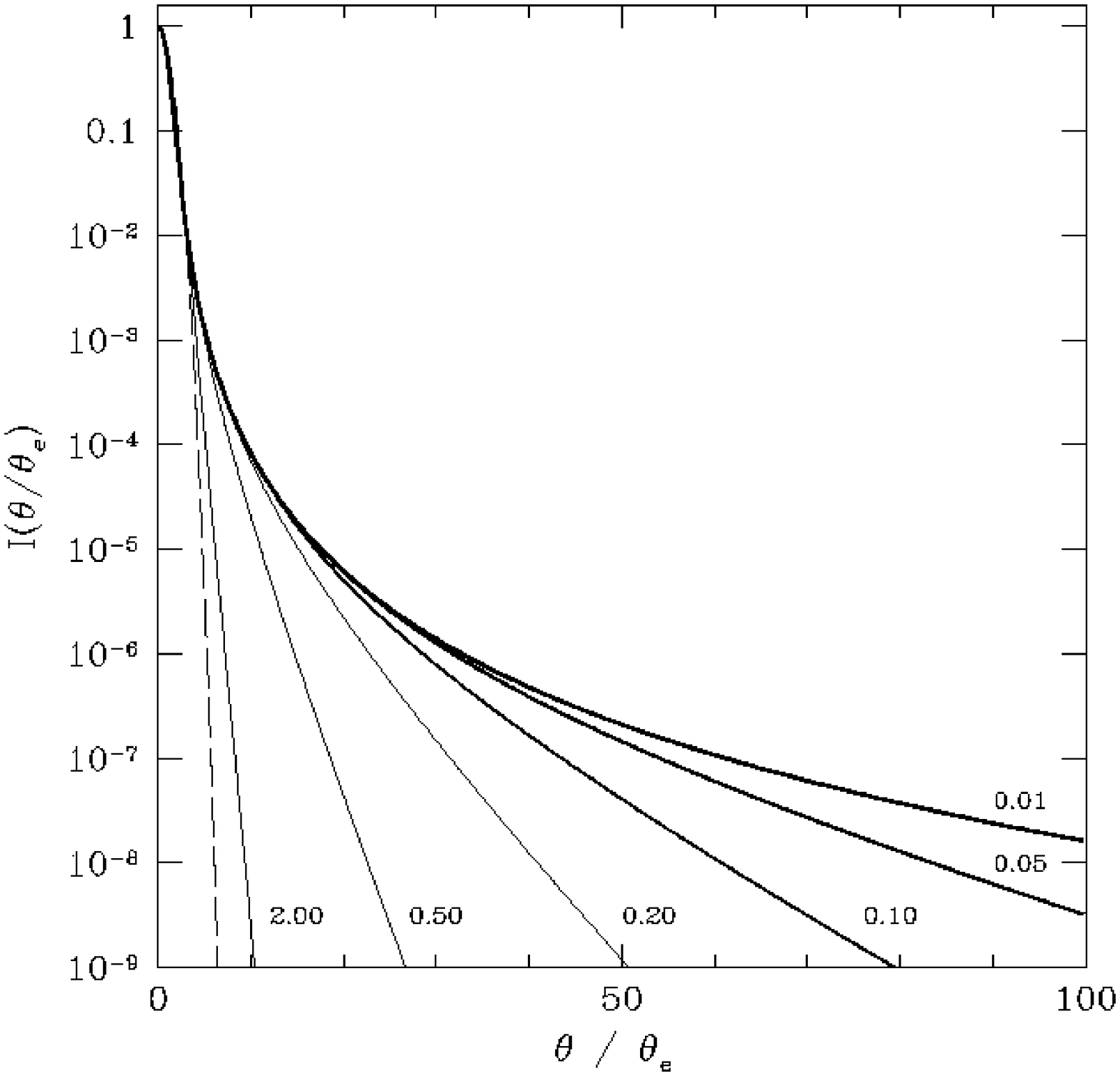}
\end{center}
\figcaption[]{
\it Left: \rm Slices through circularly symmetric images associated with
phase structure functions $\sfphi \propto b^{\alpha}$ for
$\alpha =$ 1,1.67,1.95 and 2 (thickest to thinnest lines and as labelled). 
The $\alpha = 2$ case corresponds to an image that is Gaussian in form.
\it Right: \rm Slices through circularly symmetric images for a Kolmogorov 
wavenumber spectrum with different fractional inner scales, $\zeta$,
as defined in the text and as labelled for each curve.  
The dashed curve shows a Gaussian image with the same e-folding width as the
Kolmogorov curves.
\label{fig:imageset} }
\end{figure}

\begin{figure}[htb]
\epsscale{0.4}
%\plotone{fpqlist_pow.ps}
\plotone{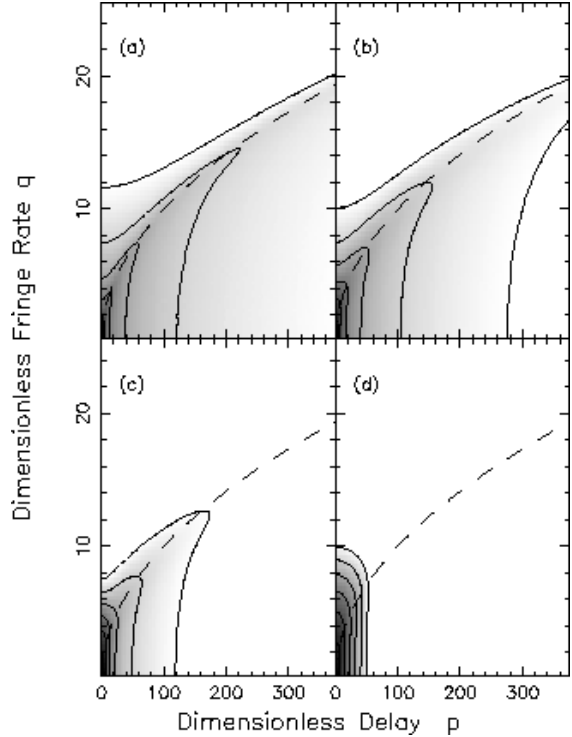}
\figcaption[]{
Secondary spectra corresponding to the image cases shown in 
the left panel of Figure~\ref{fig:imageset}. 
(a) $\alpha= 1.0$;
(b) $\alpha = 1.67$;
(c) $\alpha = 1.95$;
(d) $\alpha = 2.0$.
The contour separation is one decade and the outermost contour is
6 decades from the peak of the secondary spectrum.
The dashed line shows the parabola, $p = q^2$.
\label{fig:powfpqset} }
\end{figure}

First we consider phase structure functions of the form
$\sfphi(b) \propto b^{\alpha}$.   The corresponding wavenumber spectra
$\propto ({\rm wavenumber})^{-\beta}$ with indices
$\beta = \alpha +2$ for $\beta < 4$ (e.g. Rickett 1990). 
Figure~\ref{fig:imageset} (left-hand panel)
 shows one-dimensional slices through the
images associated with four different values of $\alpha$, including
$\alpha = 2$, which yields a Gaussian-shaped image.  The corresponding
secondary spectra are shown in Figure~\ref{fig:powfpqset}.  Arcs are most
prevalent for the smallest value of $\alpha$, become less so for larger
$\alpha$, and are nonexistent for $\alpha = 2$. 
Thus the observation of arcs in pulsar secondary spectra 
rules out the underlying
image having the form of a symmetric Gaussian function, 
and so puts an upper limit on an inner scale in the medium.
%The bulging of the contours for the $\alpha = 2$ case are the only
%residual hint of the arc effect.    

%bjr1
Arcs for all of the more extended images, 
including the $\alpha = 1.95$ case,
appear to be due to the interference
of the central ``core'' ($\psi < 1$)
with the weak ``halo'' ($\psi \gg 1$) evident in 
Figure \ref{fig:imageset} (both panels).  
This interpretation is confirmed by the work of Codona \etal\ (1986).
Their Eq.~(43) gives the cross spectrum of scintillations between
two frequencies in the limit of large wavenumbers in strong
scintillation.  When transformed,  the resulting
secondary spectrum is exactly the form of an unscattered core 
interfering with the scattered angular spectrum.

The halo brightness at asymptotically large angles 
%bjr  exponent is -(\alpha+2)
scales as $\psi^{-(\alpha+2)}$.  We can use this
in Eq.~\ref{eq:fpqg1} with the undeviated core 
as the point component ($\thvec_p=0$) 
to find $\fpq(p,q)$ far from the origin:
%Only through a detailed quantitative
%analysis can the value of $\alpha$ be inferred from the shape of the
%secondary spectrum. 
%As discussed in the previous section 
%a delta function at the origin would give infinite power density
%on the arc itself and nothing outside it.  
%For power-law structure functions  with $\alpha<2$, 
%the finite width of the image core
%prevents the divergence of the arc amplitude.
%bjr2
% much larger than angles that contribute significantly 
% to the integrand is 
\be
\fpq(p,q) \propto p^{-(\alpha+2)/2} \vert p - q^2\vert^{-1/2} 
\quad\quad  p>q^2 .
\label{eq:fpq-scaling}
\ee
Along the $p$-axis, $\fpq(p,0) \propto p^{-(\alpha+3)/2}$. 
%bjr several errors in the exponent corrected here.

%\paragraph*{Kolmogorov Spectra with an Inner Scale}
\subsubsection{Kolmogorov Spectra with an Inner Scale}

A realistic medium is expected to have a smallest (``inner'') scale in
its density fluctuations.  Depending on its size, the inner scale 
may be evident in the properties of 
pulsar scintillations and angular broadening.
Angular broadening measurements, in particular, have been used to
place constraints on the inner scale for heavily scattered lines of
sight (Moran et al. 1990;  Molnar et al. 1995; 
Wilkinson et al. 1994;  Spangler \& Gwinn 1990).
For scintillations, the inner scale can  
alter the strength of parabolic arcs, thus providing an important method
for constraining the inner scale for lines of sight with 
scattering measures much smaller than those on which angular broadening
measurements have been made. 
We consider an inner scale, $\linner$, that cuts off a 
Kolmogorov spectrum and give computed results
in terms of the normalized inner scale
$\zeta = \linner / s_0$.
The structure function scales asymptotically as  
$\sfphi \propto b^2$ below $\linner$ and 
$\propto b^{5/3}$ above.

%write the phase structure function in the ad-hoc form,
%\be
%\sfphi(\eta) =  \left [
%		\frac{1+\zeta^2}{1 + (\zeta/\eta)^2} 
%		\right]^{1/6} \eta^{5/3}. 
%\ee
%The normalized inner scale is approximately
%$\zeta = b_{\rm b} / s_0$, 
%The structure function scales asymptotically as  
%$\sfphi \propto b^2$ and $b^{5/3}$, respectively.
%(These scalings don't account for the flattening to zero slope of the
%structure function that must occur on very large baselines.) 

\begin{figure}[htb]
\epsscale{0.4}
%\plotone{fpqlist_kol.ps}
\plotone{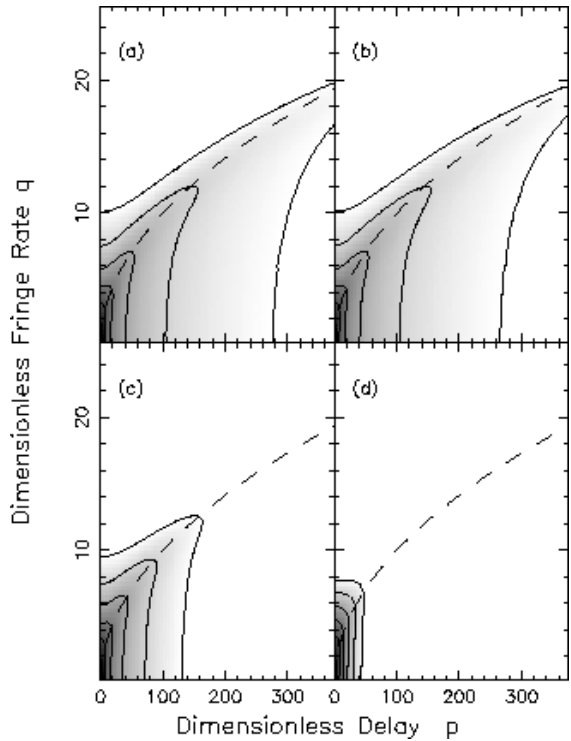}
\figcaption[]{
Secondary spectra corresponding to some of the image cases shown in 
the right panel of Figure~\ref{fig:imageset}. 
(a) $\zeta = 0.01$;
(b) $\zeta = 0.1$;
(c) $\zeta = 0.5$;
(d) $\zeta = 2.0$.
The contour separation is one decade and the outermost contour is
6 decades from the peak of the secondary spectrum.
The dashed line shows the parabola, $p = q^2$.
\label{fig:kolfpqset} }
\end{figure}

In the right panel of Figure~\ref{fig:imageset}, 
we show one-dimensional images
for six values of $\zeta$.   For $\zeta \ll 1$,  the inner scale is
negligible and the image falls off relatively slowly at large
$\psi$, showing the extended halo
$\propto \psi^{-11/3}$, as for the Kolmogorov spectrum  with no inner scale.
For $\zeta = 2$, the image falls off
similarly to the shown Gaussian form and thus does not have an extended halo.
Secondary spectra are shown in Figure~\ref{fig:kolfpqset} for four values
of inner scale.   As expected, the arcs are strongest for the case
of negligible inner scale and become progressively dimmer and truncated as
$\zeta$ increases and the image tends toward a Gaussian form.
The appearance of strong arcs in measured data indicates that the 
the inner scale must be much less than the 
%the Fresnel scale, or
%$\linner \ll 10^6$ km, and that it is much smaller than the 
diffraction scale, or $\linner \ll 10^4$ km,
for lines of sight to nearby pulsars,
such as those illustrated in 
Figures~\ref{fig:dynsec_examples}-\ref{fig:1133_variety}. 

%\paragraph*{Anisotropic Kolmogorov Spectrum}
\subsubsection{Anisotropic Kolmogorov Spectrum}

\begin{figure}[htb]
\epsscale{0.4}
%\plotone{4kolmo0.33.ps}
\plotone{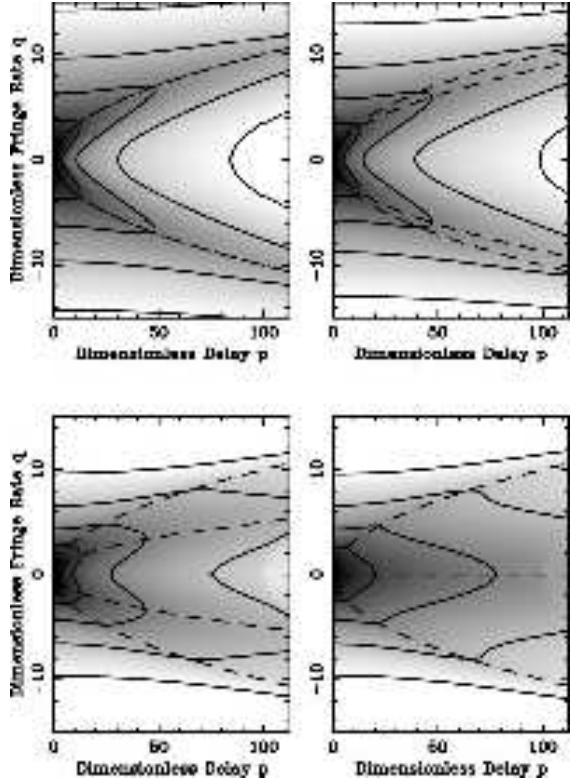}
\figcaption[]{Secondary spectra computed from 
the brightness distribution in the diffractive limit due to an
anisotropic Kolmogorov phase spectrum. The axial ratio is 3:1 with four
angles of orientation. In the upper left panel, the
brightness distribution  is extended parallel 
to the velocity; upper right at 30 $\deg$; lower left at 60 $\deg$; 
and
lower right at 90 $\deg$. Contours are at decade intervals. 
The solid curves are $p=q^2$ and $p=(q \sec\phi)^2$, corresponding
to arcs for idealized circular and linear brightness
functions.
\label{fig:aniso-kol} }
\end{figure}

Figure \ref{fig:aniso-kol} shows the results for an {\it anisotropic} 
Kolmogorov spectrum with a 3:1 axial ratio ($R = 1/3$)  for different
orientation angles of the image with respect to the velocity.
In the upper left panel the scattered 
image is elongated parallel to the velocity, and the arc is substantially
enhanced with a deep valley along the $p$-axis. 
As the orientation tends toward normal with respect to the velocity in
the other panels, the arc diminishes and essentially disappears.
The deep valley in the parallel case
occurs because $q={\theta_x}_2 - {\theta_x}_1=0$ 
along the $p$-axis, so $p = {\theta_y^2}_2 - {\theta_y^2}_1$, in which case
the secondary spectrum falls steeply like
the brightness distribution along its narrow dimension. 
By comparison, along the arc itself, the secondary spectrum probes
the wide dimension and thus receives greater weight.
Taking the asymptotic brightness at large angles,
$B \propto (R \psi_x^2 + \psi_y^2/R)^{-11/6}$ for
interference with the undeviated core
in Eq.~\ref{eq:fpqg1} we obtain: 
\be
\fpq(p,q) \propto  [q^2(R-1/R) + p/R]^{-11/6} [p-q^2]^{-1/2}  \quad\quad  p>q^2.
\label{eq:kolmo-aniso}
\ee
Along the $p$-axis the amplitude of the
secondary spectrum in Eq.~\ref{eq:kolmo-aniso} 
tends asymptotically to $\propto p^{-7/3}$.
%This puts the form of the ``valley'' and the ``edge-brightening'' factor
%quantitatively.

\section{Scattering Theory}
\label{sec:scattphys}

\subsection{Derivation of the Secondary Spectrum} 
\label{sec:spec2-theory} 

%bjr1  I changed all of this section quite a lot and moved 
% Jim's discussion of averaging into it
%
In prior sections we related $\fnu$ and $\ft$
to the angles of scattering through 
Eq.~\ref{eq:fnu} and \ref{eq:ft}.  The intensity of the
scattered waves at each angle was represented by
a scattered brightness function $B(\thvec)$
with little discussion of the scattering physics
that relates $B(\thvec)$ to the properties of the ISM.

In discussing the scattering physics, we must
consider the relevant averaging interval. 
Ensemble average results can be found in the limits 
of strong and weak scintillation, as we describe in
Appendices \ref{sec:app-strong} and \ref{sec:app-weak}.
In both cases the analysis depends on the assumption 
of normal statistics for the scattered field
and the use of the Fresnel-Kirchoff diffraction integral 
to obtain a relation with the appropriate ensemble average
$B(\thvec)$.  However, for strong scattering 
we adopt the common procedure of
approximating the observed spectrum by the ensemble 
average in the diffractive limit (after removal of the slow
changes in pulse arrival time due to changing 
dispersion measure).  In both cases $B(\thvec)$
is a smooth function of angle (except for the unscattered 
component in weak scintillation). 

%bjr1
An instantaneous angular spectrum is a single 
realization of a random process. 
Thus it will exhibit ``speckle'', i.e. the 
components of the angular spectrum are statistically 
independent having an exponential
distribution. Goodman and Narayan (1989) called
this the {\em snapshot image}, which is  obtained over times 
shorter than one diffractive scintillation time 
($\sim$minutes).  However, we average the secondary
spectrum over $\sim 1$ hour, which includes
many diffractive scintillation timescales
but is short compared to the time scale of refractive interstellar
scintillation (RISS). 
Thus we need an image obtained
from averaging over many diffractive times,
%for which we use the term {\em seeing image}, 
which Goodman and Narayan refer to as the short-term {\em average image}.  
The short-term average image has some residual speckle
but is less deeply modulated than the snapshot
image.
An approximate understanding of
the effect of speckle in the image on the secondary spectrum
is obtained from \S\ref{sec:pt+ext}, by considering
a single speckle as a point component that interferes
with an extended component that represents the rest
of the image.  Thus each speckle can give rise to part of
an arclet and multiple speckles will make an
assembly of intersecting forward and reverse arclets.
%bjr2

%The secondary spectrum may be derived in several ways.
%by formally calculating the wavefield from a thin screen using
%the, Fourier transforming over
%{\em finite} intervals in the appropriate variables, 
%and then taking an ensemble average of
%the squared magnitude of the Fourier transform.   This approach
%is aided by assuming that the thin screen's phase fluctuations have
%Gaussian statistics. In Appendix \ref{sec:app-weak}

While analytical ensemble average results for the 
secondary spectrum can be written in the asymptotic limits of 
weak and strong scattering, one must resort to
simulation for reliable results in the intermediate conditions
that are typical of many observations and in cases where
a short-term average image is appropriate.  
We use the method and code described by Coles 
et al. (1995) and analyze the resulting diffraction pattern
in the same manner as for the observations.
We first create a phase screen at a single frequency
whose wavenumber spectrum follows a specified
(e.g. the Kolmogorov) form. 
We then propagate a plane wave through 
the screen and compute the resulting 
complex field over a plane at a distance $z$ beyond the screen.
The intensity is saved along a slice through the plane of observation,
giving a simulated time series at one frequency.
The frequency is stepped, scaling the screen phase as for
a plasma by the reciprocal of frequency, and the process is repeated.
The assembly of such slices models a dynamic spectrum, which is
then subject to the same secondary spectral analysis
as used in the observations. 

The results, while calculated for
a plane wave source, can be mapped to a point source using 
the well-known scaling transformations for the screen geometry
(i.e. $z\to \dse$, as defined in Eq.~\ref{eq:defdse}).
While the screen geometry is idealized, the simulation
uses an electromagnetic calculation and
properly accounts for dispersive refraction, diffraction 
and interference. Furthermore the finite size of
the region simulated can approximate the finite integration
time in the observations.  A scalar field is used 
since the angles of scattering are extremely
small and magnetionic effects are negligible in 
this context (c.f. Simonetti et al. 1984).

\subsection{Interstellar Conditions Responsible for Arcs}
\label{sec:ISconditions}

In \S \ref{sec:theory} we found two 
conditions that emphasize the arcs.  The
first is the interference between an undeviated wave and 
scattered waves, and the second is the enhancement of arcs
when the waves are scattered preferentially 
parallel to the scintillation velocity.
We now discuss how these conditions might occur.

%\paragraph*{Core-Halo Conditions}
\subsubsection{Core-Halo Conditions.}

The basic arc, $p=q^2$, is formed by interference
of a core of undeviated or weakly deviated waves  with widely 
 scattered waves.
There are two circumstances where
a significant fraction of the total flux density 
can come from near $\thvec = 0$.  
One is in strong scintillation, when
the ``central core'' of the scattered brightness 
interferes with a wider low level halo,
as discussed in \S\ref{sec:powerlaw-sfn} for
a Kolmogorov density spectrum and other 
power-law forms with $\beta < 4$ and neglible inner scale.
In this case the core of the scattered brightness 
is not a point source, merely much more compact than the
far-out halo radiation.  Thus the arc is a smeared version
of the form in Eq.~\ref{eq:fpqU}, which has no power outside
the basic arcline. Examples were shown in figures
\ref{fig:powfpqset} and \ref{fig:kolfpqset}.

The other case is when the scattering 
strength is low enough that a fraction of 
the flux density is not broadened significantly in angle. 
The requirements on such an ``unscattered core'' 
are simply that the differential delay over 
the core must not exceed, say, a quarter of the wave 
period. Then the core is simply that portion of the brightness 
distribution in the first Fresnel zone, $\rf$,  
and the electric-field components in the core sum coherently. 
Thus the core component can be significant even if the overall 
scattered brightness function is much wider than 
the first Fresnel zone.

The relative strength of the core to the remainder of the brightness 
distribution can be defined in term of the ``strength of scintillation''. 
We define this as $m_{\rm B}$, the normalized rms intensity 
(scintillation index) under the Born approximation. It can be written 
in terms of the wave structure function $D_{\phi}(r)$, as 
$\mbsq = 0.77 D_{\phi}(\rf)$ with a simple Kolmogorov spectrum, 
where $D_{\phi}(r)=(r/s_0)^{5/3}$, $s_0$ is the field 
coherence scale and $\rf =\sqrt{\dse/k}$
and $k=2\pi/\lambda$.  Here $\thd=1/ks_0$ defines 
the scattering angle. This gives
$\mbsq = 0.77 (\rf/s_0)^{5/3}$; thus it is also related to 
the fractional diffractive bandwidth $\sim (s_0/\rf)^2$.
Note that a distinction should be made between 
strength of {\em scattering} and strength of {\em scintillation}.
For radio frequencies in the ISM we always expect
that the overall rms phase perturbations will be 
very large compared to one radian, albeit over relatively 
large scales, which corresponds to
strong {\em scattering}.  In contrast strong {\em scintillation}
corresponds to a change in phase of more than one
radian across a Fresnel scale. 

\begin{figure}
\begin{center}
\begin{tabular}{c}
\epsscale{0.25}
\includegraphics[angle=0,width = 12cm]{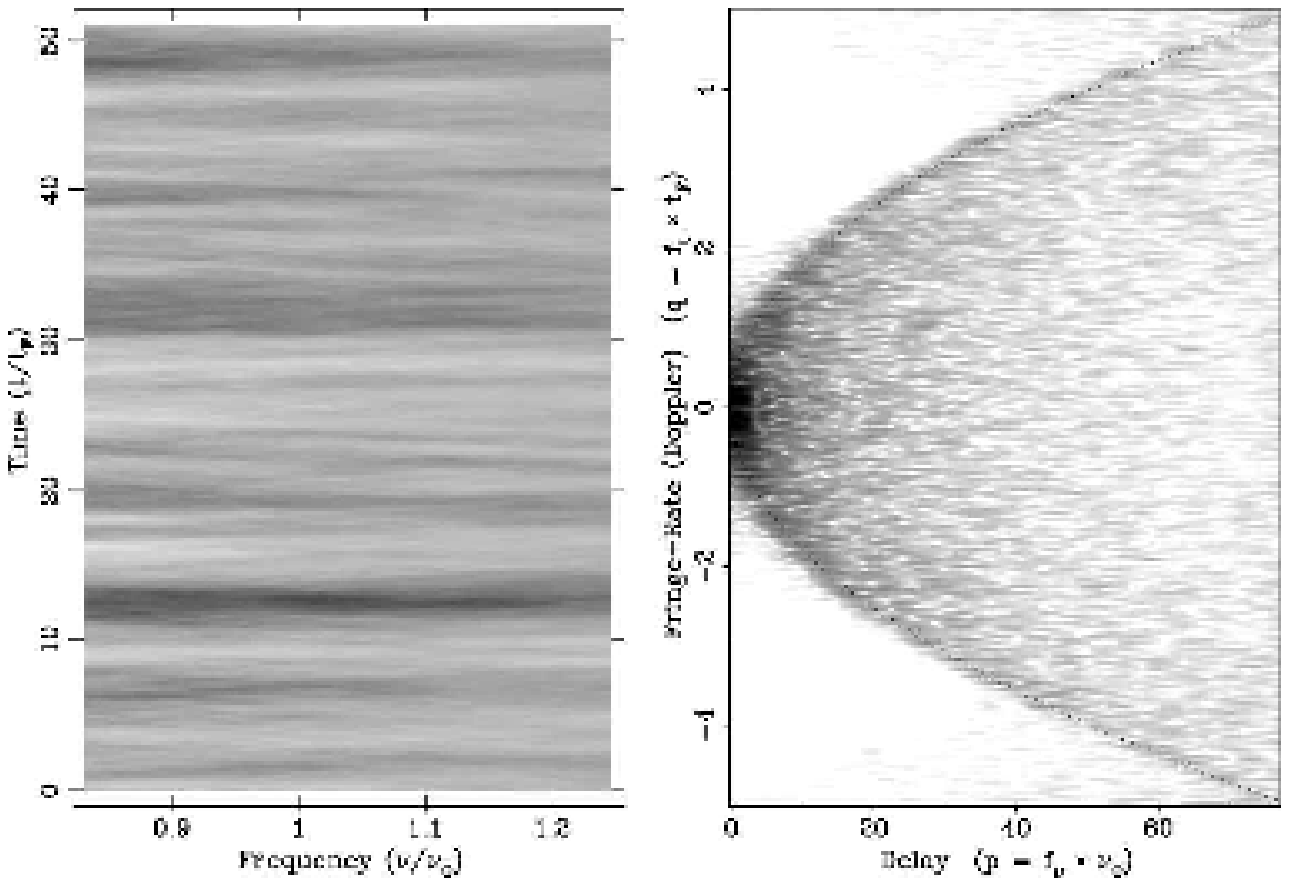} \\
\includegraphics[angle=0,width = 12cm]{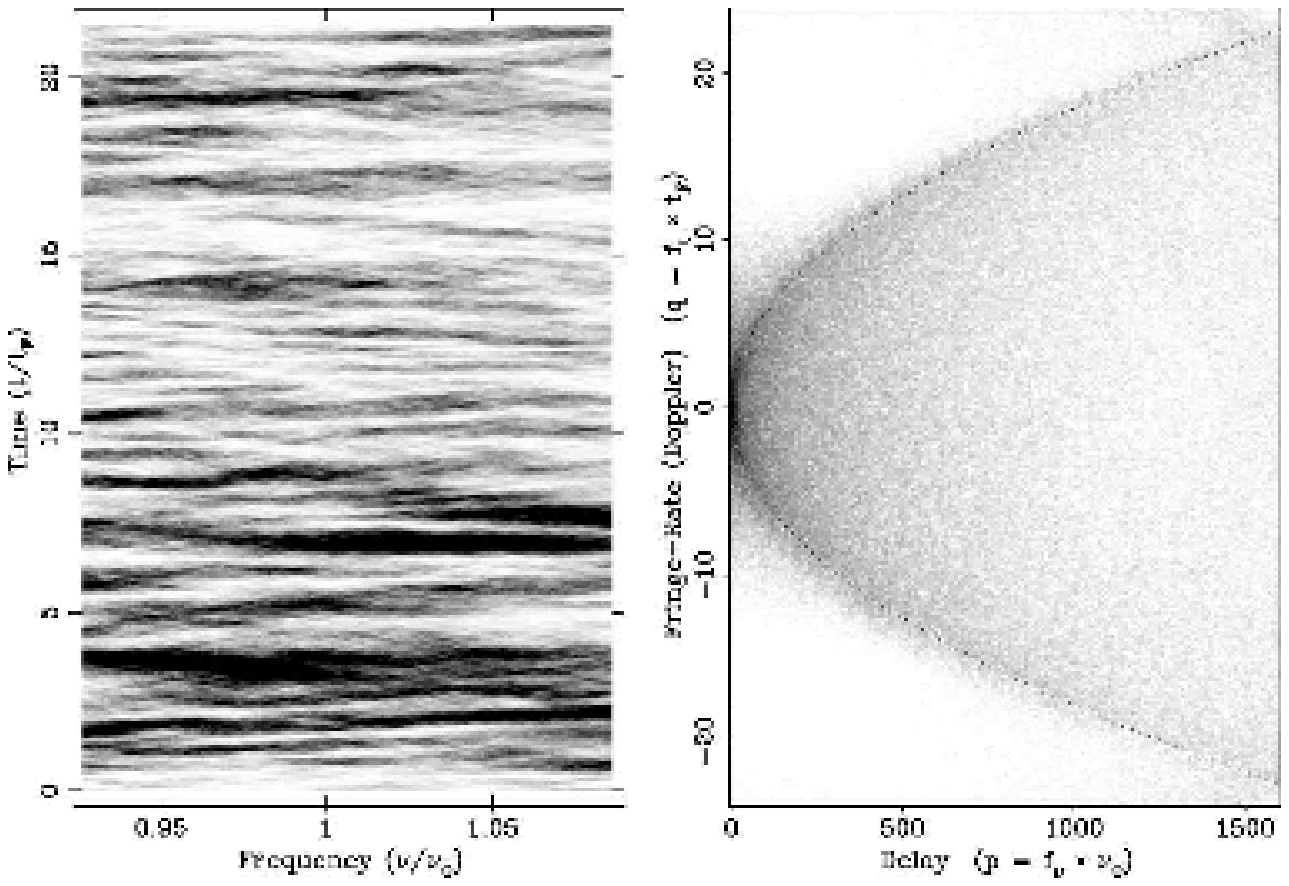}
\end{tabular}
\end{center}
\figcaption[]{ Primary and secondary spectra computed 
from simulations of plane wave incident
on a phase screen with an isotropic Kolmogorov spectrum. 
\it Top: \rm $\mbsq=0.1$; \it  Bottom: \rm $\mbsq=10$.
On the left the primary dynamic spectrum has a grey scale
proportional to the intensity from zero (white) to 
three times the mean (black). On the right the secondary 
spectrum has a grey scale proportional to the logarithm
of the spectral density over a dynamic range of 5.5 decades (top) and
6.5 decades (bottom).
Note the steep decrease outside the theoretical arcline
($p=q^2$, shown dotted).
\label{fig:simkoliso} 
}
\end{figure}

In weak scintillation (but strong scattering) 
the fraction of the flux density
in the core is $\sim (\theta_{\rm F}/\thd)^2 = 
(s_0/\rf)^2 \sim m_{\rm B}^{-2.4}$. 
Figure \ref{fig:simkoliso} (top panels) shows a simulation with
$\mbsq=0.1$, in which $\fpq$ drops steeply outside the 
arc $p=q^2$.  
The analysis of Appendix \ref{sec:app-weak}
shows that in weak scintillation $\fpq$ should
be analyzed as a function of wavelength 
rather than frequency because
 the arcs are more clearly delineated when the 
simulations are Fourier analyzed versus wavelength.
We have also simulated weak scintillation in a nondispersive phase
screen and find similar arcs.
In strong scintillation there is a small fractional 
range of frequencies and the distinction between wavelength and
frequency becomes unimportant. 

Figure \ref{fig:simkoliso} shows that the secondary spectrum 
in weak scintillation has a particularly well defined arc 
defined by a sharp outer edge. This is because it is 
dominated by the interference between scattered waves
and the unscattered core, as described by Eq.~(\ref{eq:fpqg1}) 
with a point component at the origin. We can ignore the mutual 
interference between scattered waves
since they are much weaker than the core.  
The result is that the secondary spectrum is analogous
to a hologram in which the unscattered core serves as
the reference beam.  The secondary spectrum
can be inverted to recover the scattered brightness function
using Eq.~(\ref{eq:fpqg1}), which is the analog of viewing a 
holographic image.  The inversion, however, is
not complete in that there is an ambiguity in the equation
between positive and negative values of $\theta_y$.
Nevertheless, the technique opens the
prospect of mapping a two-dimensional
scattered image from observations with a single dish
at a resolution of milliarcseconds.

As the scintillation strength increases the core 
becomes less prominent, and the parabolic arc loses contrast.
However, we can detect arcs at very low levels 
and they are readily seen at large values of $\mbsq$.
Medium strong scintillation is shown
in Figure \ref{fig:simkoliso} (bottom panels), which is a simulation
of a screen with an isotropic Kolmogorov spectrum 
with $\mbsq = 10$ for which $s_0 = 0.22\rf$.
The results compare well with both
the ensemble average, strong 
scintillation computations for a Kolmogorov screen 
in \S\ref{sec:ismcases} and with several 
of the observations shown in Figure \ref{fig:dynsec_examples}.
We note that several of the observations 
in \S\ref{sec:obs} show sharp-edged and symmetric
parabolic arcs, which appear to be more common for
low or intermediate strength of scintillation (as characterized
by the apparent fractional diffractive bandwidth).
This fits with our postulate that the arcs become 
sharper as the strength of scintillation decreases.
In weak scintillation the Born approximation applies, in which case
two or more arcs can be caused by two or more screens
separated along the line of sight.  We have confirmed this by 
simulating waves passing through several screens, each of which yields
a separate arc with curvature as expected for an unscattered wave
incident on each screen.  This is presumably the explanation for the 
multiple arcs seen in Figure~\ref{fig:1133_variety} for pulsar
B1133+16.

In terms of normalized variables $p,q$ the 
half-power widths of $\fpq$ are approximately unity
in strong scattering
and yet we can see the arcs out to $q\gg 1$. As we noted
earlier this corresponds to scattering angles well
above the diffractive angle $\thd$, and
so probes scales much smaller than those probed
by normal analysis of diffractive ISS.
Our simulations confirm that with an inner scale 
in the density spectrum having a wavenumber cutoff
$\kappa_{\rm inner}$, the arc will be reduced beyond 
where $q \sim \kappa_{\rm inner} s_0$. 
However, to detect such a cut-off observationally 
requires very high sensitivity and a large dynamic 
range in the secondary spectrum.  

%\paragraph*{Anisotropy, Sub-Arcs and Speckle.}

%\input arctheory.v11.5.2.2.tex
%Revised 5.2.2    drafted June 3, 2004  BJR
%Changes by DRS 2004-06-15:
% 	denoted  %drs
%   * corrected figure references now that simkoliso and 
%     simkolaniso are distinct figs
%   * discussion of N_x and N_y in terms of anisotropy R
%  * N.B.  Made a reference to possible ESE connection
%     new refs:  Fiedler et al. 1987; Romani, Blandford, & Cordes 1987

%\paragraph*{Anisotropy, Sub-Arcs and Speckle.}
\subsubsection{Anisotropy, Arclets and Image Substructure}

\begin{figure}
\begin{center}
\begin{tabular}{c}
\epsscale{0.25}
%\includegraphics[angle=-90,width = 12cm]{submbsq10-2048-512-3.ps} \\
%\includegraphics[angle=-90,width = 12cm]{mbsq10-phgrad3.ps}
%\includegraphics[angle=-90,width = 12cm]{f11a.eps} \\
%\includegraphics[angle=-90,width = 12cm]{f11b.eps}
%\vspace{-0.3truein} 
\includegraphics[angle=-0,width = 12cm]{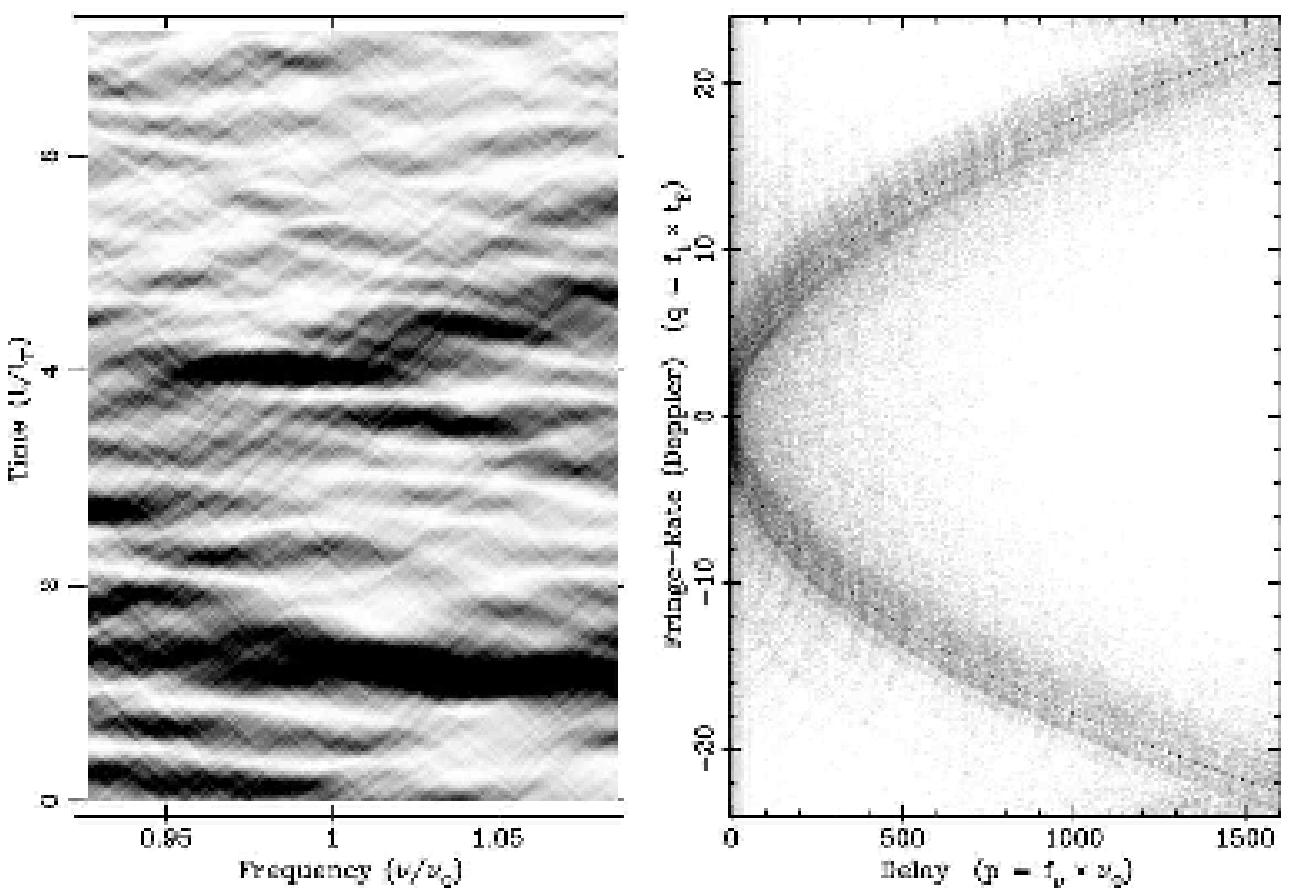} \\
\includegraphics[angle=-0,width = 12cm]{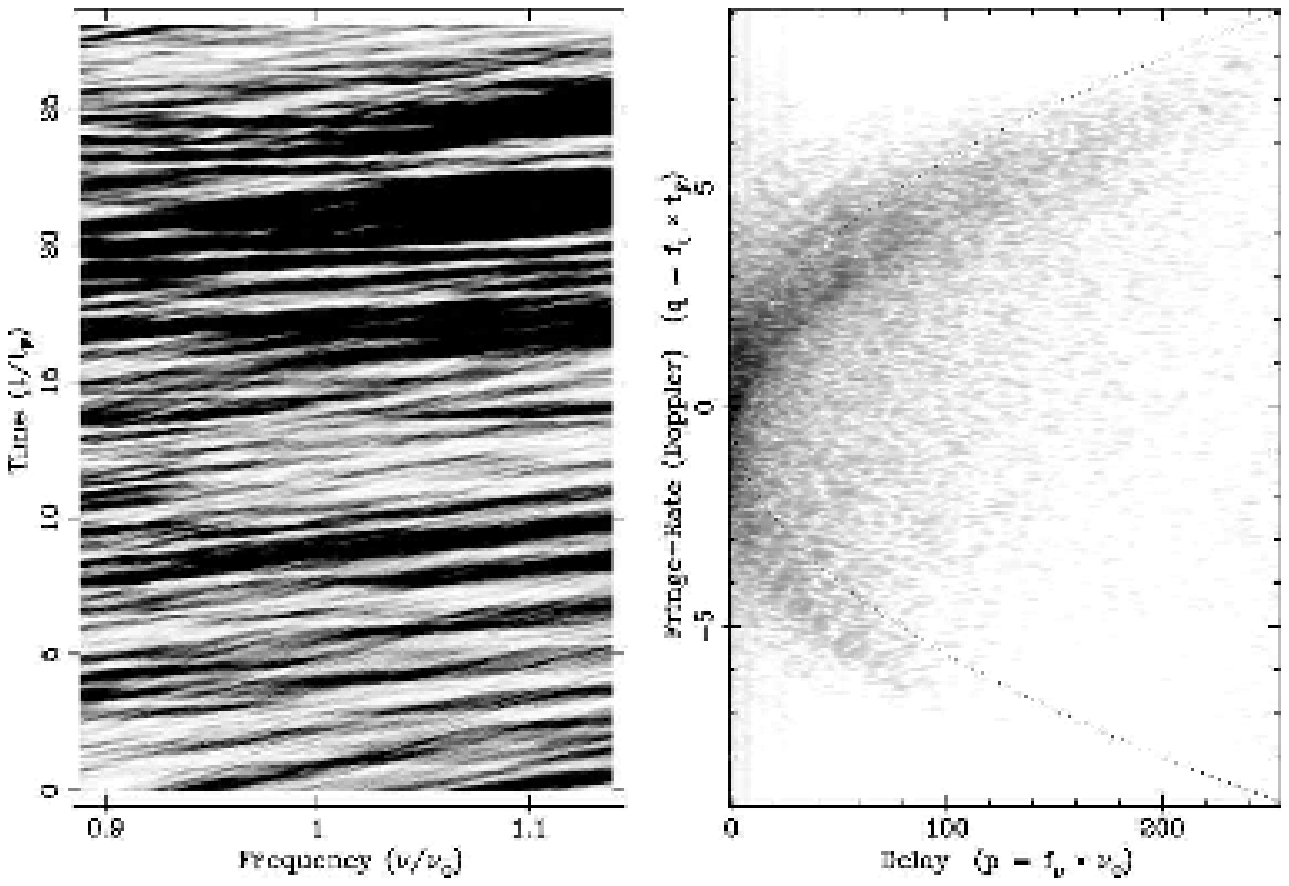}
\end{tabular}
\end{center}
\figcaption[]{Simulations of a phase screen as in 
the bottom panel of Figure~\ref{fig:simkoliso} with $\mbsq = 10$. 
\it Top: \rm Anisotropic scattering (axial ratio 4:1 
spectrum elongated along the velocity direction). 
The result is higher contrast in the arc, 
which is seen to consist of nearly parallel arclets 
and is crossed by equally fine scale reverse arcs.
\it Bottom: \rm Isotropic scattering with a linear
refractive phase gradient of 2 radians per diffractive scale.
The result is a shift in the parabola and an asymmetry
in its strength.
\label{fig:simkolaniso} 
}
\end{figure}

The enhancement in arc contrast when the scattering
is extended along the direction of the effective
velocity was discussed in \S\ref{sec:theory}. The
enhancement is confirmed by comparing the 
$\mbsq = 10$ simulations in Figures~\ref{fig:simkoliso} 
and \ref{fig:simkolaniso}.
%drs anisotropic $\mbsq = 10$ simulation in the upper panels of Figure 
%\ref{fig:simkoliso}  with the isotropic case 
%in the  lower panels. 
% Figure \ref{fig:simkolaniso}.  

In the anisotropic case the arc consists 
of finely-spaced parallel arcs 
%drs arcs used here instead of arclets
which are in turn crossed by reverse 
arclets, somewhat reminiscent of the
observations.  The discussion of \S\ref{sec:pt+ext}
suggests that these are caused by the mutual 
interference of substructure in the angular spectrum 
$B(\thvec)$. 
%drs parentheses removed around B(\thvec)
The question is, what causes the substructure?
One possibility is that it is stochastic speckle-like substructure
in the image of a scatter-broadened source.
Alternatively, there could be discrete
features in $B(\thvec)$, 
(``multiple images'') caused by particular 
structures in the ISM phase. 

In our simulations, the screen is stochastic
with a Kolmogorov spectrum and so it should have 
speckle but no deterministic multiple images.
%drs The size of speckles in an image is set 
%by the angular resolution of the telescope.
%However, it is not clear what sets the effective
%angular resolution in our observations.
The diffractive scale, $s_0$, is approximately the 
size of a stationary phase point on the screen and, also,
approximately the size of a constructive maximum
at the observer plane.
There are, thus, 
$\sim (\rf/s_0)^4$ across the 2d scattering disk that contribute
to an individual point at the observer plane.
This large number of speckles ($\sim 400$ in these simulations)
has the effect of averaging out details in the secondary spectrum,
particularly in the isotropic scattering case. 

%drs discussion of N_x and N_y in terms of anisotropy R
The effect of anisotropy in the coherence scale is to broaden the image
along one axis with respect to the other.
An image elongated by a ratio of $R = \theta_x / \theta_y \approx
s_{0,y} / s_{0,x} > 1$ would have $N_x \approx( \rf/s_{0,x})^2$ speckes in the $x$-direction but only 
$N_y \approx( \rf/ R s_{0,x})^2$ in the $y$-direction.
In short, the image would break 
into a line of many elliptical speckles (elongated
in $y$) distributed predominantly along $x$. In comparing 
the anisotropic and isotropic simulations 
(upper Figure \ref{fig:simkolaniso} and lower Figure \ref{fig:simkoliso}),
it appears that the arclets are more visible 
under anisotropy. We suggest that the changed substructure
in the image is responsible.  Substructure may consist of short-lived
speckles or longer-lived multiple subimages whose relative contributions
depend on the properties of the scattering medium.
In the simulations the arclets appear to
be independent from one realization to the 
next, as expected for a speckle phenomenon.
However, in some of the observations arclets
persist for as long as a month
(see \S\ref{sec:obs}) and exhibit a higher contrast than
in the simulation.  
%drs added "persist for a month" and "higher contrast"
These
require long-lived multiple images in
the angular spectrum and so imply 
quasi-deterministic structures in
the ISM plasma. 

In summary we conclude that the occasional isolated arclets
(as in Figure \ref{fig:0834_multifreq})
must be caused by fine \em substructure \rm in $B(\thvec)$.
Whereas some of these may be stochastic as 
in the speckles of a scattered image, 
the long lived arclets require
the existence of discrete features in 
the angular spectrum, which cannot be part 
of a Kolmogorov spectrum.  These might be
more evidence for discrete structures in the medium at 
%drs had been greater than 1 AU
scales $\gtrsim 1$~AU, that have been invoked
to explain fringing episodes 
(e.g. Hewish, Wolszczan \& Graham 1985;
Cordes \& Wolszczan 1986; Wolszczan \& Cordes 1987; 
 Rickett, Lyne \& Gupta 1997) and,
 %drs added a reference to ESE's
 possibly, extreme scattering events (ESEs)
 (Fiedler \etal 1987; Romani, Blandford \& Cordes 1987)
 in quasar light curves.  

%\paragraph*{Asymmetry in the Secondary Spectrum.}
\subsubsection{Asymmetry in the Secondary Spectrum}

As summarized in \S\ref{sec:obs} the observed arcs
are sometimes asymmetric as well as
exhibiting reverse arclets (e.g. Figure 
\ref{fig:0834_multifreq}).  
%In \S\ref{sec:pt+ext} we discussed the mutual interference
%of a point and extended components. Equation \ref{eq:fpqg1} 
%shows that when the point or extended component
%is offset in $\theta_x$, 
A scattered brightness that is asymmetric in $\theta_x$
can cause $\fpq$ to become asymmetric in $q$.
We do not expect a true ensemble average brightness
to be asymmetric, but the existence of large
scale gradients in phase will refractively shift
the seeing image, with the frequency dependence
of a plasma.  

Simulations can be used to study refractive effects, as
in Figure \ref{fig:simkolaniso} (bottom panels), which shows $\fpq$
for a screen with an isotropic Kolmogorov spectrum
 and a linear phase gradient.
The phase gradient causes sloping features in the primary
dynamic spectrum and asymmetry in the secondary spctrum.  
The apex of the parabolic arc is shifted to negative 
$q$ value and becomes brighter for positive $q$.
Thus we suggest refraction as the explanation of the 
occasional asymmetry observed in the arcs.  
Relatively large phase gradients are needed
to give as much asymmetry as is sometimes observed.
For example, a gradient of 2 radians per $s_0$,
which shifts the scattering disc by about its diameter,
was included in the simulation of Figure \ref{fig:simkolaniso} to
exemplify image wandering.
In a stationary random medium with a Kolmogorov 
spectrum, such large shifts can occur depending on the 
outer scale,  but they should vary only over times
long compared with the refractive scintillation timescale.
This is a prediction that can be tested.

In considering the fringe frequencies in Appendix \ref{sec:appendixa}
we only included the geometric contribution
to the net phase and excluded the plasma term.
We have redone the analysis to include a plasma term
with a large scale gradient and curvature in the screen phase
added to the small scale variations which cause
diffractive scattering. We do not give the details here
and present only the result and a summary of the issues involved.

We analyzed the case where there are large scale
phase terms following the plasma dispersion law, 
that shift and stretch the unscattered ray, as in the 
analysis of Cordes, Pidwerbetsky and Lovelace (1986). 
In the absence of diffractive scattering
these refractive terms create a shifted stationary phase 
point (i.e. ray center) at $\thvecbar$ and
a weak modulation in amplitude due to
focusing or defocusing over an elliptical region.
With the diffractive scattering also included we find that 
the fringe frequency $q$ is unaffected by refraction
but the delay $p$ becomes:
%\subsection{Refractive Modification of Images and Secondary Spectra}
%\label{sec:refraction}
%
%In Appendix~\ref{sec:appendixa}, we give the image that results when
%the diffraction-only image is distorted by a component of the phase screen
%that refracts  as well as diffracts incident radiation.  The case we
%analyze is one where the length scales for the refraction are much larger
%than those causing diffraction.   
%Refraction generally displaces the image, 
%increases the ellipticity, and modifies the integrated intensity. 
%We therefore expect the secondary spectrum to be modified  
%in accord with properties that we have identified through previous examples.
%In Appendix~\ref{sec:appendixa} we show that the conjugate frequency
%$q$ is unaffected by refraction whereas $p$ takes on the form
\be
p &=&         \left(\thvec_2+\thvecbar\right)^2  - 
              \left(\thvec_1+\thvecbar\right)^2 \nonumber \\ 
% I rewrote the first part to show 
&-& 
		\frac{\lambda_0}{2\pi} \dse
                  \left[
                  (\thvec_2-\thvecbar)\cdot\Cmatrix\cdot(\thvec_2 - \thvecbar)
                 -(\thvec_1-\thvecbar)\cdot\Cmatrix\cdot(\thvec_1 - \thvecbar)
                  \right],
\ee
%where $\thvecbar$ is the displacement of the image and 
where $\Cmatrix$ is a $2\times2$ matrix that 
describes the quadratic dependence of the
refracting phase from the image center.
%  (see Appendix).   It may be seen easily that
%the first term, $2\thvecbar\cdot (\thvec_2 - \thvec_1)$, results
%when a shifted image $B(\thvec-\thvecbar)$ is inserted into
%Equation~\ref{eq:fpq1} and a change of variables is made to
%unshift the image, thus modifying $p$ but not $q$.   The terms involving
%$\Cmatrix$ correspond to rescaling and rotating the image. 

A related question  is whether the shift in image position due
to a phase gradient also shifts the position of
minimum delay, in the fashion one might expect from Fermat's
principle that a ray path is one of minimum delay.
However, since Fermat's minimum delay is a phase delay
and our variable $\fnu$ or $p$ is a group delay, its minimum
position is shifted in the opposite direction by a 
plasma phase gradient. This is shown by the first two
terms, i.e.\ $(\thvec_2+\thvecbar)^2$ 
which goes to zero at $\thvec_2=-\thvecbar$.
However, we do not pursue this result here, since the formulation of
\S\ref{sec:theory} does not include the frequency dependence of the 
scattered brightness function, which would be required for a full
analysis of the frequency dependence of the plasma scattering.
It is thus a topic for future theoretical study, 
but meanwhile the simulations provide the insight that indeed
plasma refraction can cause pronounced asymmetry in the arcs.

%\paragraph*{Arcs from Extended Sources}
\subsubsection{Arcs from Extended Sources}

Like other scintillation phenomena, arcs will be suppressed
if the source is extended.  In particular, the lengths of arcs
depend on the transverse spatial extent over which the scattering screen
is illuminated by coherent radiation from the source.  For a source of finite
angular size $\theta_{\rm ss}$ as viewed at the screen,
the incident wave field is spatially incoherent on scales larger than 
%$b_{\rm ss} \sim \lambda/(\pi \theta_{\rm ss})$.    
$b_{\rm ss} \sim \lambda/(\theta_{\rm ss})$.   
An arc measured at a particular fringe rate ($\ft$) represents
the interference fringes of waves separated
by a baseline $\bvec = (D-\ds)(\thvec_2 - \thvec_1)$ 
at the screen, corresponding to a fringe rate
$\ft = \bvec\cdot \vperpvec / \lambda \dse$.
The fringes are visible only if the wave field is coherent
over $\bvec$ and are otherwise suppressed.
Thus arcs in the secondary spectrum $S_2$ will be cut off for fringe rates
$\ft\gtrsim f_{\rm t, sou}$, where
\be
%f_{\rm t, sou} = \frac{s\vperp}{\pi \dse \theta_{\rm sou}}.
f_{\rm t, sou} = \frac{s\vperp}{\dse \theta_{\rm sou}}.
\label{eq:ft-sou}
\ee
%In writing Eq.~\ref{eq:ft-sou},
Here
we distinguish between the source size $\theta_{\rm sou}$
viewed at the observer and its size viewed at the screen,
$\theta_{\rm ss}=\theta_{\rm sou}/s$. For ISS of a
distant extragalactic source, the factor $s$ approaches unity
but can be much smaller for a Galactic pulsar.   
Eq.~\ref{eq:ft-sou} indicates
that longer arcs are expected for more compact sources, larger
effective velocities, and scattering screens nearer to the observer.
Equivalently, using $\fnu = a\ft^2$, the arc length can be measured 
along the $\fnu$ axis with a 
corresponding cut-off, 
\be
\fnu = 
%\frac{s}{\pi\nu_0}  
\frac{s}{\nu_0}  
\left (\frac{\theta_{\rm Fr}} {\theta_{\rm sou}} \right)^2
\; ,
\label{eq:fnu-sou}
\ee
where  $\theta_{\rm Fr}=[kD(1-s)]^{-1/2}$ is the effective
Fresnel angle. 

It is useful to measure the arc's extent in $\ft$ in terms of the
characteristic DISS time scale, $\dtd \sim s_0 / \vperp$ and to relate
the characteristic diffraction scale $s_0$ to the isoplanatic angular
scale, $\theta_{\rm iso} \sim s_0 / (D - \ds)$.  
The product of maximum fringe rate and DISS time scale is then
\be
%f_{\rm t, sou} \dtd \approx \frac{s_0}{\pi \dse \theta_{\rm ss}} 
f_{\rm t, sou} \dtd \approx \frac{s_0}{\dse \theta_{\rm ss}} 
	\approx \frac{\theta_{\rm iso}}{\theta_{\rm sou}}.
\label{eq:ftarclength}
\ee
The isoplanatic angular scale defines the source size
that will quench DISS by $\sim 50$\%.  
We note that Eq.~\ref{eq:ftarclength} is consistent with
the extended source result for scintillation modulations
(Salpeter 1967).
Thus the length of the arc
along the $\ft$ axis in units of the reciprocal DISS time scale is a direct
measure of the ratio of isoplanatic angular scale to source size.
The long arcs seen therefore demonstrate that emission regions are much
smaller than the isoplanatic scale, which is typically 
$\lesssim 10^{-6}$ arcsec for measurements of dynamic spectra.

The theoretical analysis under weak scintillation conditions
is given in Appendix D.2, where it is seen that the squared
visibility function provides a cut-off to the point-source 
secondary spectrum $S_2$.  A remarkable result from this analysis
is that measurements of $S_2$ of an extended source
can be used, in principle, to estimate
the  squared visibility function of the source in two dimensions -- if 
the underlying secondary spectrum,
$S_2$, for a point source is already known.
%When the source is circularly symmetric,
%the cut-off in fringe rate is consistent with Eq.~\ref{eq:ft-sou},
%which applies to a normal scintillation spectrum at frequency $\ft$,
%due to the visibility cut-off (e.g. Salpeter, 1967).

The effects of an extended source can also be analyzed
in asymptotic strong scintillation. This requires the
same type of analysis as for the frequency decorrelation
in scintillation of an extended source.  
Chashei and Shishov (1976) gave the result for a medium 
modeled by a square law structure function of phase. 
Codona \etal\ (1986) gave results for screens with 
a power law spectrum of phase in both weak and strong 
scintillation.  We have used their analysis
to obtain an expression for the secondary spectrum in
the strong scattering limit. The result is that 
the source visibility function appears as an additional factor 
$|V|^2$ inside the brightness integral of 
Eq.~(\ref{eq:fpq1}) (with arguments depending on $p,q$ and other
quantities).

It is clear that the detection and measurement of arcs from pulsars
can put constraints on the size of their emitting regions.
This is intimately related to estimating 
source structure from their occasional episodes of 
interstellar fringing 
(e.g. Cordes and Wolszczan 1986; Wolszczan and Cordes 1987;
Smirnova et al. 1996; Gupta, Bhat and Rao 1999).
These observers detected changes in the phase of the fringes
versus pulsar longitude, and so constrained any spatial offset
in the emitting region as the star rotates.  They
essentially measured the phase of the ``cross secondary spectrum'' 
between the ISS at different longitudes,
at a particular $\fnu,\ft$. Clearly one could extend this 
to study the phase along an arc in $\fnu,\ft$.
Such studies require high signal-to-noise ratio 
data with time and frequency samplings that resolve scintillations
in dynamic spectra, which can be obtained on a few pulsars with 
the Arecibo and Green Bank Telescopes.  The future Square Kilometer
Array with $\sim 20$ times the sensitivity of Arecibo would allow
routine measurements on large samples of pulsars.

ISS has been seen in quasars and active galactic nuclei
(sometimes referred to as intraday variations),
but few observations have had sufficient frequency coverage to
consider the dynamic spectrum and test for arcs.  However,
spectral observations have been reported for the quasars 
J1819+385 (de Bruyn and Macquart, in preparation) and B0405-385 
(Kedziora Chudczer et al., in preparation). A preliminary
analysis of the data from J1819+385 by Rickett et al. (in preparation)
showed no detectable arc, from which they set a lower limit on
the source size.  New observations over a
wide well-sampled range of frequencies will allow
better use of this technique.

%bjr2  

%\input arctheory.v11.sec6.tex

%drs changes as of 2004-06-16 -- just a few
\section{Discussion and Conclusions}
\label{sec:discuss}

It is evident that we have only begun to explain the
detailed structures in the parabolic arcs observed
in pulsar scintillation. However, it is also clear
that the basic phenomenon can be understood from a
remarkably simple model of small angle scattering from a thin
phase-changing screen, and does not depend on the dispersive
nature of the refractive index in the screen.
Interference fringes between pairs of scattered waves
lie at the heart of the phenomenon. 
%While in the asymptotic
%strong scattering limit the interference is between
%components of the scattered brightness function, in
%the seeing image regime it is expressed as interference
%between spherical waves scattered from two points in the screen.
The $\fnu$ coordinate of the secondary spectrum is readily
interpreted as the differential group delay between the
two interfering waves, and the coordinate $\ft$ is interpreted
as their fringe rate or, equivalently, 
the differential Doppler frequency, which is proportional
to the difference in angles of scattering projected along
the direction of the scintillation velocity.

We have developed the theory by modeling the interference
for an arbitrary angular distribution of scattering.
We have given the ensemble average secondary
spectrum in asymptotic strong and weak 
scintillation, and we have used a full phase screen simulation
to test the results under weak and intermediate strength of
scintillation.  The results are mutually consistent.

A simple parabolic arc with apex at the origin of the $\fnu,\ft$
plane arises most simply in weak scintillation as the interference
between a scattered and an ``unscattered'' wave. The secondary
spectrum is then of second order in the scattered field
%and is given by an integral over the intensity
%cross-spectrum, which when evaluated 
and maps to the
two-dimensional wavenumber spectrum of the screen phase,
though with an ambiguity in the sign of the wavenumber perpendicular
to the velocity.  Remarkably, this gives a way to estimate
two-dimensional structure in the scattering medium
from observations at a single antenna,
in a fashion that is analogous to holographic reconstruction.

In strong scattering the parabolic arcs become 
less distinct since the interference between
two scattered waves has to be summed over all possible angles
of scattering, making it a fourth order quantity 
in the scattered field. Nevertheless, the arc
remains visible when the scattered brightenss has a compact
core and a ``halo'' of low level scattering at relatively
large angles. Media with shallow power-law wavenumber spectra 
(including the Kolmogorov spectrum) have
such extended halos, and the detection of arcs provides
a powerful probe of structures ten or more times smaller
than those probed by normal interstellar scintillation
and can thus be used to test for an inner scale
cut-off in the interstellar density spectrum.

The prominence of arcs depends on the isotropy
of the scattering medium as well as on the slope and inner scale
 of its wavenumber spectrum. 
Arcs become more prominent when the scattering is anisotropic
and enhanced preferentially along the scintillation
velocity. However, 
%drs clarifcation - simulations
in simulations, prominent arcs are seen over quite 
a wide range of angles about this orientation.
Scattering that is enhanced parallel to
the scintillation velocity
corresponds to \it spatial \rm structures that are
elongated transverse to the  velocity vector.
Thus the common detection of arcs may provide
evidence for anisotropy in the interstellar plasma,
and with careful modeling observations should yield estimates
for the associated axial ratios.  

There are several details of the observed arcs for which
%drs word order change:  we only have --> we have only
we have only tentative explanations. We can understand
the existence of discrete reverse arcs as due to 
discrete peaks in the scattered brightness interfering with
an extended halo. Such isolated peaks are to be expected in
short term integrations due to speckle in the scattered image.
However, observations with only a few isolated reverse arcs
%drs - addition here (next line)
-- and, particularly, arclets that persist for days to weeks --
imply only a few discrete peaks in the scattered image, while
normal speckle is expected to give multiple bright points with
a much higher filling factor. This is a topic for further 
investigation.

Another observational detail is that on some occasions
the arc power distribution is highly asymmetrical in
fringe frequency.  This can only be caused by asymmetry 
in the scattered brightness relative
to the velocity direction.  Our proposed explanation
is that it is due to large scale gradients in the
medium that cause the image to be refractively shifted.
The simulations demonstrate that this explanation is feasible, 
but considerable more work needs to be done to 
interpret what conditions in the ISM are implied
by the unusual asymmetric arcs.  

Our theoretical analysis is based on a thin screen
%drs comma added in next line
model, and future theoretical work is needed on the arc phenomena
with multiple screens (such as might cause the multiple arcs
in Figure \ref{fig:1133_variety}) and with an extended scattering
medium. While the extension to an extended medium or
%drs straight forward --> straightforward
multiple screens is relatively straightforward in 
weak scintillation, it is more difficult in strong
scintillation.  Adding the effect of a source
with a finite diameter is also important since pulsar emission regions
may have finite size and 
the detection of arcs from quasars provides the prospect
of a more powerful probe of their angular structure
than from simple analysis of their scintillation 
light curves.  In addition to these extensions of our analysis,
future work will include 
a detailed study of the inverted arclet phenomenon,
exploiting the arc phenomenon to determine the anisotropy and inner scale 
  of scattering irregularities;
and using the multiple arc phenomenon to aid modeling of the local
interstellar medium, for which the weak scattering regime is especially 
relevant.

\acknowledgments
We acknowledge helpful discussions with D. Melrose and M. Walker. 
DRS wishes to thank Oberlin students 
H.~Barnor, D.~Berwick, A.~Hill, N.~Hinkel,
D.~Reeves, and A.~Webber for assistance in the 
preparation of this work.  
This work was supported by the National Science Foundation through 
grants to Cornell (AST 9819931 and 0206036), 
Oberlin (AST 0098561) and 
UCSD (AST 9988398).
This work was also supported by the National Astronomy and Ionosphere
Center, which operates the Arecibo Observatory under a cooperative
agreement with the NSF.
The Australia Telescope National Facility provided
hospitality for DRS during preparation of this paper.

\clearpage

\appendix

% original:
%\def\bfs{{\bf x}}
%\def\bsigma{{ \mbox{\boldmath $\sigma$} }}
%\def\nuo{{\nu_0}}
% new: s->x   \sigma -> \Delta x
% bjr  to conform with appendix A new: s->\rvec
%  s_o to \rvec
%  s_p to \rsvec
%  sprime to \rvecp
%   \sigma -> \Delta x  ->  \Delta \rvec
\def\bfs{{\bf x}}
\def\delrvec{{ \mbox{\boldmath $\Delta$} \rvec}}
\def\nuo{{\nu_0}}
% bjr   also change field 'f' to \varepsilon
%  \nu to \Delta \nu   using \nu as the radio frequency 
\def\delnu{{\Delta\nu}}
\def\f{{\varepsilon}}

%\centerline{\bf APPENDIX}
%\input arctheory.v11.appA.tex

\section{Fringe Frequencies from a Plasma Screen}
\label{sec:appendixa}

\def\f{{\varepsilon}}

Consider the following thin-screen  geometry:
a point source at $(\rsvec, 0)$,
a thin screen in the plane $(\rvecp, D_s)$ and 
an observer at $(\rvec, D)$,
where $\rsvec, \rvecp$ and $\rvec$ are two dimensional vectors.
The screen changes the phase of incident
waves and thus diffracts and refracts radiation from the source.
%Our goal here is to discuss the influence
%of refraction on the relations given by
%equations (\ref{eq:fnu} and \ref{eq:ft}).

The Kirchoff diffraction integral  
(KDI) gives the wave field at $\rvec$ as
\be
\f(\nu, \rvec) = \left(i\lambda\dse\right)^{-1}
	\int d\rvecp\, e^{i\Phi(\nu, \rvec)},
\label{eq:kdi}
\ee
using the effective distance $\dse$ 
as defined in Eq.~\ref{eq:defdse} and
where $\Phi = \Phi_g + \phi_d $ is the 
sum of the geometric phase,
\be
\Phi_{\rm g} = \frac{k}{2} 
	 \left [
		\ds^{-1} \vert \rvecp-\rsvec\vert^2
		+
		(\dt)^{-1} \vert \rvec-\rvecp\vert^2
	 \right ],
\label{eq:phi_g}
\ee
a diffractive phase $\phi_d(\rvecp)$ that scatters radiation.
% and a refractive phase $\phi_r$ that distorts images 
% and may introduce multiple images.
%bjr1
%The diffractive phase represents the influence
%of the small scales in the medium, while  the refractive
%phase represents the large scales (larger than
%the size of the scattering disk of the diffractive phase).
%bjr2
Frequency scalings are 
$\Phi_{\rm g}\propto \nu$ and $\phi_{d} \propto \nu^{-1}$.

The secondary spectrum is the distribution of
conjugate frequencies $\fnu,\ft$ produced by all pairs of
exit points from the screen.  
Consider the relative  phase, 
$\Delta\Phi = \Phi_2 - \Phi_1$,
between two components of the radiation
field that exit the phase screen at two different points,
$\rvecp_{1,2}$, that correspond to deviation 
angles as viewed by the observer, ${\thvec}_{1,2} = \rvecp_{1,2} / (D-\ds) $.
%\Phi(\nu,t,\rvec) = 
%\Phi(\nu,t,\rvec, \rvecp_2) -\Phi(\nu,t,\rvec, \rvecp_1)$.
The combined radiation from the two points will oscillate
as a function of time, frequency, and spatial location.  For fixed
location on axis ($\rvec=0$) and using the effective velocity
(Eq.~\ref{eq:veff}) to map spatial offsets at the screen to time, 
we can expand $\Delta\Phi$ in time and frequency offsets:
\be
\Delta\Phi(\nu,t) = 
	\Delta\Phi(\nubar,0) + 2\pi [\fnuab (\nu-\nubar) + \ftab t],
\ee
where 
(using $\partial_t \equiv \partial/\partial_t$, etc.) 
the fringe frequencies are
\be
\ftab &=& (2\pi)^{-1} \partial_t \Delta\Phi(\nu,t) \\
\fnuab &=& (2\pi)^{-1} \partial_{\nu} \Delta\Phi(\nu,t).
\ee
Here we use only the geometric phase to
calculate the fringe frequencies. The result is given
by equations (\ref{eq:fnu} and \ref{eq:ft}) in terms of the
two apparent angles ${\thvec}_{1,2}$ and the effective
velocity $\vperpvec$ (equation \ref{eq:veff}).
%bjr-start
Since the delay is defined in terms of the frequency
derivative of the phase, $\fnu$ is the difference
in the \it group \rm delay. While the distinction is
unimportant for the geometric phase, it
makes a difference for the dispersive plasma contributions.
When the analysis is done including the
derivatives in these plasma terms,
the equation for $\fnu$ is modified but there is
%bjr-end
no change in the equation for $\ft$, as mentioned
at the end of \S\ref{sec:ISconditions}.

In appendices \ref{sec:app-strong} and \ref{sec:app-weak}
a derivation is given of the secondary spectrum
in the limits of strong and weak scintillation,
respectively.  In the strong limit we find that
it is given explicitly by the double integral over
the observed angles of scattering in
equation (\ref{eq:fpq1}). We note that the
integrand is the scattered brightness function, defined as
a spectrum of plane waves. In contrast the discussion given
above in terms of the KDI is based on spherical waves
emanating from
%bjr-start
\it points \rm in the screen.
While one cannot equate the apparent
angular position of points on the screen,
${\thvec}_{1,2} = \rvecp_{1,2} / (D-\ds) $,
to the angles of arrival of plane waves components
in the angular spectrum, one can obtain identical equations
for $\fnu$ and $\ft$ by considering
an integral over plane waves emanating from a screen,
which is illuminated by a point source. The method is
similar to the KDI analysis above, except that
one expands the propagation phase of each
plane wave component as a function of frequency and time.
Its derivatives give $\fnu$ and $\ft$, precisely, as in equations
(\ref{eq:fnu} and \ref{eq:ft}).
%bjr-end

\section{Finite Resolution and Numerical Issues for the
Secondary Spectrum}
\label{sec:app-num}

Empirically, the secondary spectrum is estimated over
a finite total bandwidth $B$ and integration time $T$ with 
finite resolution in frequency and time.
These in turn set finite resolutions in $\fnu$ and $\ft$
and so in $p$ and $q$.
%and also determine the frequency range over which the
%scattered brightness function is to be evaluated.
The integral expressions for the secondary spectrum
such as Eq.~(\ref{eq:fpq1}) diverge 
at the origin of the p-q plane since they ignore resolution effects.

%With finite resolutions, $\Delta p, \Delta q$,
%the integrand in Equation~\ref{eq:fpqxy} is replaced by 
%the integral
%\be
%\frac{1}{\Delta z} \int_{\Delta z/2}^{\Delta z/2} dz\, 
%	\fthvec\left( X_{-}+z, \sqrt{\qabs}\,(x+y)\right)
%        \fthvec\left( X_{+}+z, \sqrt{\qabs}\,(x-y)\right),
%\ee 
%where $\Delta z = \Delta p / 2\qabs$. 
%For small-enough $\Delta z$,
%the integral can be approximated as the integrand evaluated
%at $z=0$. For this to be true we need $\Delta z \ll W_{\rm x}$, 
%where $W_{\rm x}$ is the width of $\fthvec$ in the $x$ direction.
%For normalized angles, $W_{\rm x} \equiv 1$, so the approximation
%requires $\qabs \gg \Delta p / 2$.   Therefore if Equation~\ref{eq:fpqxy}
%is used to calculate a model secondary spectrum,  non-zero values of
%$q$ are needed that satisfy this inequality.  
%In numerical examples discussed below, we 
%use $\qabs \gtrsim 0.05$ when the angles are normalized.   

Finite resolution in $p$ can be included by replacing the Dirac delta
functions in Eq.~(\ref{eq:fpq1}) by rectangular functions of unit
area whose limiting forms are delta functions. 
%\be
%\delta(p-\pbar) \to \Delta(p-\pbar,\Delta p),
%\ee 
%using a function $\Delta(x,\Delta x)$ defined as a unit area
%rectangle function of full width $\Delta x$,
%\be
%\Delta(x,\Delta x) \equiv $
%	\frac{1}{\Delta x} \Pi \left ( \frac{x}{\Delta x} \right ), 
%\ee
%where $\Pi(x) = 1$ for $\vert x \vert \le \Delta x/2$ and zero otherwise.
%
Then performing the integrations over $d\thvec_2$ yields a form 
for $\fpq$, 
\be
\fpq(p,q) \approx 
	\sum_{\pm}
	\int d\thvec_1
        \fthvec(\thvec_1) 
%\left [ 
%\fthvec({\theta_1}_x+q, \sqrt{U})
%+ \fthvec({\theta_1}_x+q, -\sqrt{U})
%\right ]
	\fthvec({\theta_1}_x+q, \pm\sqrt{U})
	H'(U-\Delta p/2) 
    \frac{\sqrt{U + \Delta p / 2} - \sqrt{U - \Delta p/2}}
     {\Delta p}  \; ,
\label{eq:fpqfinite}
\ee
where $U = ( {\theta_1}_y^2 + p - q^2 - 2q{\theta_1}_x) $
and the summation is over the two ideal solutions
$\theta_{2y}=\pm\sqrt{U}$, and we have ignored the variation in $B$ 
over the range $\Delta p$ near each solution;
$H'$ is a modified unit step function with a transition width $\Delta
p$.  As $\Delta p\to 0$, the factors 
involving $\Delta p$ tend toward a delta
function.  For finite $\Delta p$, however, $\fpq(0,0)$ remains finite.

In terms of the bandwidth $B$ and time span $T$,
the resolutions in $p$ and $q$ 
are (when angles are normalized by the diffraction angle, $\thd$)
\be
%\Delta p &=& \frac{2\pi\dnud}{C_1\Delta\nu} = \frac{2\pi\epsilon}{N_{\nu}}, \\
\Delta p &=& \frac{1}{B \tau_d} = \frac{2\pi\epsilon}{N_{\nu}}, \\
\Delta q &=& \frac{2\pi\dtd}{T}= \frac{2\pi\epsilon}{N_{\rm t}},
\ee
where $N_{\rm t}$ and $N_{\nu}$ are, respectively, the number
of distinct `scintles' along the time and frequency axes, respectively
(here, $\epsilon\sim 0.2$ is a constant that quantifies the filling
factor of scintles in the dynamic spectrum; e.g. Cordes \& Lazio 2002).
The resolutions of $p$ and $q$ therefore are determined by how many scintles
are contained in the dynamic spectrum, which in turn determine the 
statistical robustness (through $N^{-1/2}$ effects) 
of any analyses of a particular dynamic spectrum.
For typical dynamic spectra, $N_{\rm t}$ and $N_{\nu}$ are each 
$\gtrsim 10$, so $\Delta p$ and $\Delta q$ are each $\lesssim 0.1$.
Observationally, the individual channel bandwidth and sampling time
are also important since they determine the Nyquist points in $p$ and $q$.

In computing the secondary spectrum from the integral
in Eq.~(\ref{eq:fpqU}) one also needs to resolve the
(integrable) singularity $U^{-1/2}$.
This can be achieved by changing variables to
$s_{x,y} = ({\theta_2}_{x,y} + {\theta_1}_{x,y})/2\sqrt{\qabs}$ and
$d_{x,y} = ({\theta_2}_{x,y} - {\theta_1}_{x,y})/2\sqrt{\qabs}$, and
integrating over the delta functions. Then letting
$x = s_y$ and $y = -d_y$ we obtain
\be
\fpq(p,q) &=&
\int\int dx\,dy\,
        \fthvec\left( X_{-}, \sqrt{\qabs}\,(x+y)\right)
        \fthvec\left( X_{+}, \sqrt{\qabs}\,(x-y)\right), 
\label{eq:fpqxy}
\\ 
X_{\pm} &=& \frac{ p\pm q^2}{2q} + 2\, {\rm sgn}(q) x y, 
\ee 
where ${\rm sgn}(q)$ is the sign of $q$.
Note that symmetry of $\fpq$ upon letting $p\to - p$ and
$q\to -q$ is demonstrated by also letting $y \to -y$.

\section{The Secondary Spectrum in the Strong Scattering Limit}
\label{sec:app-strong}

%%bjr1  change again to make the notation the same as
% Jim's APPENDIX A
%

Here we present secondary spectrum
expected in the asymptotic limit case of strong scintillations
from a single phase screen.
The intensity from a point source recorded at 
position $\rvec$ and frequency $\nu$ 
is the squared magnitude of the phasor $\f$ for the electric field:
$I(\rvec,\nu) = | \f(\rvec,\nu) |^2$, where the dependence on
source position $\rsvec$ is suppressed.
From the dynamic spectrum of a pulsar we can define the correlation of
intensity versus offsets in both space and frequency. 
After subtracting the mean,
$\Delta I = I - <I>$, the correlation function is:
\be
R_{\Delta I}(\delrvec,\delnu) =\, < \Delta I(\rvec,\nu) \Delta 
I(\rvec +\delrvec,\nu+\delnu) > .
\label{eq:rdi}
\ee
Under asymptotic conditions of  strong scintillation the phasor $\f$ becomes
a Gaussian random variable with zero mean and random phase,
the real and imaginary parts of $\f$ are uncorrelated, and the
fourth moment can be expanded in products of 
second moments.  It follows that:
\be
R_{\Delta I}(\delrvec,\delnu) = |\Gamma(\rvec,\delrvec,\nu,\nu+\delnu) |^2 \; ,
\label{eq:RdelI}
\ee
where
\be
\Gamma(\rvec,\delrvec,\nu,\nu+\delnu) = \,  
< \f(\rvec,\nu) \f^{*}(\rvec +\delrvec,\nu+\delnu) > 
\; .
\ee
When the field $\f$ is due to a point source 
scattered by a phase screen
at distance $\ds$ from the source and 
$\dds$ from the observer (with $D$ the pulsar
distance), the second moment is the  product (see LR99): 
\be
\Gamma(\rvec,\delrvec,\nu,\nu+\delnu) =  \Gamma_{\rm point} \Gamma_R \Gamma_D.
\ee
Here $\Gamma_{\rm point}$ is simply due to the spherical wave
nature of a point source, which is essentially unity for typical
pulsar observations and $\Gamma_R$ is due to the wandering of dispersive
travel times about its ensemble average  as 
the electron column density changes, $\Gamma_R = \exp(-\pi^2 
\delnu^2 \tau_R^2)$. 
$\Gamma_D$ is the diffractive second moment, which is given in terms of the
scattered angular spectrum $B(\thvec)$ at the 
radio frequency $\nu$:
\be
\Gamma_D(\delrvec,\delnu) = \int\, d\thvec  
	B(\thvec) \exp[-  2\pi i \delnu \theta^2 D (1-s)/(2 c s)] 
\exp(i k \thvec \cdot \delrvec); 
\label{eq:GammaD}
\ee
recall that $s = \ds/D$.
The phase term in the first exponential 
is proportional to the extra delay 
$\theta^2 D (1-s)/(2 c s)$
for waves arriving at the observer at an angle $\thvec$;  
this quadratic relation 
between time delay and angle of arrival gives rise to the 
quadratic features in the secondary spectrum.
In single-dish pulsar observations the spatial offset $\delrvec$ is 
sampled by a time offset $t$ times the relative 
velocity of the diffraction pattern past the observer.  
Such observations are in a short-term regime in which 
the dispersive delay is essentially constant over the integration time
and so observations are characterized by $\Gamma_D$,
the diffractive second moment at $\delrvec = \vperpvec t D/\ds$;
the distance ratio is needed since $\vperpvec$ is the effective
screen velocity.
The secondary spectrum is the double Fourier Transform
of the correlation function $R_{\Delta I}$, 
\be
S_2(\fnu,\ft) = \int dt\, \int d\delnu\,
e^{2\pi i \ft t   + 2\pi 
\fnu \delnu} 
R_{\Delta I}(\vperpvec t D/\ds,\delnu) \;.
\label{eq:PDeltaI}
\ee
In the short-term regime, 
this equation is evaluated using \ref{eq:GammaD} for $\Gamma_D$
in place of $\Gamma$ in Eq.~\ref{eq:RdelI}.  Integration over
$t$ and $\delnu$ and conversion to the scaled variables of
\S\ref{sec:theory} yields Eq.~\ref{eq:fpq1} in the main text.

\section{The Weak Scintillation Limit}
\label{sec:app-weak}

% changes in notation to conform with JMC Appendix A   bjr

We now examine the secondary spectrum in the limit of weak
scintillation due to a plasma phase screen with a power law
wavenumber spectrum. 
%bjr  I shortened the weak scint analysis by 
% referring to Codona et al 1986  for the cross(frequency) spectrum

\subsection{Point Source}

%Consider the complex field $\f(\rvec,\rsvec,\lambda)$
%at wavelength $\lambda$ observed at transverse coordinate 
%$\rvec$ at a distance $\dds$ from a 
%phase screen $\phi(\rvecp)$,
%which is illuminated by a point 
%source at transverse coordinate 
%$\rsvec$ a distance $\ds$ behind the screen. 
%In all that follows we ignore the decrease of 
%intensity as the inverse square of distance, 
%and $\f$ is normalized such that
%the average intensity $<|\f|^2>=1$. 
%We find it convenient to specify the observing 
%wavelength instead of the frequency, and use 
%$D$ for the distance from the source to the observer. 
%Using the Fresnel-Kirchoff
%diffraction integral $\f$ can be written as
%\be
%\f(\rvec,\rsvec,\lambda) = e^{-i\Phi} 
%e^{-i\phi(\bfre)} h(\bfre,\lambda,\dse)  \; . 
%\ee
%Here $\bfre= s\,\rvec  + (1-s)\,\rsvec $ is the 
%lateral coordinate where a straight line 
%from the point source to 
%the observer passes through the screen,
%$\Phi = kD + k|\rsvec-\rvec|^2/2 D$ is 
%the free-space phase term and
%\be
%h(\bfre,\lambda,\dse) = \frac{ik}{2\pi\dse} \int d^2\rvecp
%\exp[i(\phi(\bfre)-\phi(\bfre-\rvecp) )] \exp[-ikr'^2/2\dse)] \; .
%\label{eq:h-fresnel}
%\ee
%At the screen $\dse=0$ and $h=1$ and at a distance $D$ 
%we write $h=1+\delta h$. Near the screen the scintillation 
%is weak $|\delta h| \ll 1$.  

Scintillation is said to be weak when
the point-source, monochromatic  
scintillation index (rms/mean intensity) is much
less than one.  This applies near a phase screen
since  intensity fluctuations only build up as a wave 
travels beyond the screen.  In this regime the KDI (Eq.~\ref{eq:kdi})
can be approximated by a first order expansion of 
$\exp[i\phi_d(\rvecp)]$, where the screen phase at $\rvecp$ is
written as the screen phase at the observer coordinate
$\rvec$ plus the difference in phase between 
$\rvecp$ and $\rvec$.  This allows a linearization
of the problem,  even though the rms 
variation in overall screen phase may be
very large, as expected for a power law spectrum.

Various authors have described the frequency dependence of
scintillation under these conditions.
Codona et al.\ (1986), in particular, give a thorough
analysis applicable to evaluating the secondary spectrum.
They obtain expressions for the correlation of intensity fluctuations
between different observing wavelengths ($\lambda_1, \lambda_2$),
the quantity in our Eq.~\ref{eq:rdi}.
Under weak scintillation conditions the result is most simply
expressed in terms of its 2-D Fourier transform over $\kappavec$, 
the cross-spectrum of intensity fluctuations:
\be
P_{\Delta I}(\kappavec,\lambda_1,\lambda_2) =
\int d \kappavec R_{\Delta I}(\Delta \rvec,\lambda_1,\lambda_2)
\exp(i \kappavec \cdot \Delta\rvec)/(4\pi^2) \; .
\ee
Here we find it convenient to work in terms of observing wavelength
rather than frequency, because it simplifies the weak scintillation
results.

Codona et al.\ give an expansion for the cross-spectrum
applicable to low wavenumbers in their equation (27).
This is the product of the wavenumber spectrum of the screen phase
with the two-wavelength ``Fresnel filter''
and with an exponential cut-off applicable to strong refractive
scintillation, which can be ignored in weak scintillation.
Their results are given for a non-dispersive
phase screen and a plane incident wave; when converted to a plasma 
screen and the point source geometry described in previous
sections the result is:
\be
P_{\Delta I}(\kappavec,\lambda_1,\lambda_2) =
(\lambda_1 \lambda_2/\lambda_0^2) P_{\phi}(\kappavec)
2 [\cos(\kappa^2\dse\lambda_d/2\pi) - \cos(\kappa^2\dse\lambda_0/4\pi)] \; .
\label{eq:pdi-wk}
\ee
where $\dse= Ds(1-s)$, $\lambda_d = \lambda_1-\lambda_2$ and 
$\lambda_0= (\lambda_1+\lambda_2)/2$ and
$ P_{\phi}(\kappavec)$ is the wavenumber spectrum 
of screen phase at  $\lambda_0$.
Note that with $\lambda_d=0$ the difference in the two cosine functions
become the well-known $\sin^2$ Fresnel filter.

%The intensity $I(\rvec,\rsvec,\lambda)=|\f|^2=|h|^2$, which
%in weak scintillation can be approximated as:
%\be
%I(\rvec,\rsvec,\lambda)= 1 + \delta h + \delta h^* = 1 + \delta I \; ,
%\label{eq:weakI}
%\ee
%where 
%\be
%\delta I(\bfre,\lambda)= \frac{k}{2\pi \dse} \int d^2\rvecp \;
%[\phi(\bfre)- \phi(\bfre-\rvecp)]\, 2\cos(kr'^2/2\dse) \; .
%\ee
%Here we evaluate the secondary spectrum as the 
%two-dimensional transform versus difference 
%in time and wavelength instead of time and frequency:
%\be
%P_{\Delta I}(\ft,f_{\lambda}) = \int\int d\tau d\lambda_d 
%\exp[2\pi i(\ft\tau+ f_{\lambda} \lambda_d)]
%R_{\Delta I}(\tau,\lambda_1,\lambda_2)
%\label{eq:pdi-weak1}
%\ee
%where $\lambda_d = \lambda_1-\lambda_2$ and 
%$\lambda_0= (\lambda_1+\lambda_2)/2$ and
%\be
%R_{\Delta I}(\tau,\lambda_1,\lambda_2) =
%< \delta I(\bfre=\vperpvec t,\lambda_1) 
%\delta I(\bfre=\vperpvec (t+\tau),\lambda_2) >  \; .
%\ee
%The appropriate velocity is given by Eq.~(\ref{eq:veff}).
%We assume $\phi$ scales as $\lambda$ for a plasma screen and obtain:

Defining the secondary spectrum as the Fourier transform 
with respect to time difference and wavelength difference
instead of frequency difference, we obtain 
\be
S_2(f_{\lambda},\ft) = \int d\lambda_d e^{2\pi i f_{\lambda}
\lambda_d} \int d\kappavec P_{\Delta I}(\kappavec,\lambda_1,\lambda_2)
\delta(\ft-\kappavec\cdot\vperpvec/2\pi) \; ,
\ee
where we have not included the finite resolution effects
or the integration with respect to the mean 
wavelength $\lambda_0$ over the total bandwidth $B$. 
Substituting from Eq.~\ref{eq:pdi-wk} we obtain:
\be
S_2(f_{\lambda},\ft) = (\lambda_0 z)^2 \int d\lambda_d
 [1-(\lambda_d/2\lambda_0)^2] 
\exp[2\pi i f_{\lambda} \lambda_d] \times \nonumber \\
\int d\kappavec  P_{\phi}(\kappa_x,\kappa_y) 
\delta(f_t-\kappavec\cdot\vperpvec/2\pi)
\, 2[\cos(\kappa^2\dse\lambda_d/4\pi) -
\cos(\kappa^2\dse\lambda_0/2\pi)] \; .
\label{eq:pdi-weak2}
\ee

For small fractional bandwidths we 
approximate $1-(\lambda_d/2\lambda_0)^2 \sim 1$
and, taking the $x$-axis along $\vperpvec$,
the integrals can  be evaluated to give:
\be
S_2(f_{\lambda},\ft) = 
\frac{8\pi^3 H(\kappa_{yp}^2)}{\vperp \dse \kappa_{yp}} 
\left [ P_{\phi}(\kappa_x=\frac{2\pi f_t}{\vperp},\kappa_{yp})+
P_{\phi}(\kappa_x=\frac{2\pi f_t}{\vperp},-\kappa_{yp})
\right ]  \nonumber \\
- \delta(f_{\lambda}) P_{\rm w}(f_t)  \; .
\label{eq:pdi-weak3}
\ee
Here $H(u)$ is the unit step function and 
\be
\kappa_{yp} = \sqrt{8\pi^2 |f_{\lambda}|/\dse 
-(2\pi f_t/\vperp)^2}
\label{eq:kyp}
\ee
$P_{\rm w}(f_t)$ is
\be
P_{\rm w}(f_t) = \frac{4\pi}{\vperp} \int d\kappa_y
\cos\left\{\frac{\dse\lambda_0}{4\pi}
\left[\frac{2\pi^2 f_t^2}{\vperp^2}+\kappa_y^2\right]\right\} \,
P_{\phi}(\kappa_x=\frac{2\pi f_t}{\vperp},\kappa_{y})  \; .
\ee
It is closely related to the
normal weak scintillation spectrum at wavelength $\lambda_0$,
with the difference that the Fresnel filter ($\sin^2( )$)
function is replaced here by $\sin^2( )-1/2$.
Excluding the $P_w$ term in Eq.~(\ref{eq:pdi-weak3}), 
we see that $S_2$ diverges
along the parabolic curve where $\kappa_{yp}=0$, 
creating a parabolic arc, and is cut-off
by the step function outside that curve.  
With circularly symmetric scattering, $P_{\phi}$ is a function only of 
$|\kappa|^2 = 8\pi^2 |f_{\lambda}|/\dse$ and so
the dependence on $f_t$ is purely through
the known arc-enhancement factor $1/\kappa_{yp}$
using Eq.~\ref{eq:kyp}.  Thus a measurement
of $S_2(f_{\lambda,\ft})$ can be inverted
to estimate  $P_{\phi}(\kappa_x,\kappa_y)$, 
and so we have a direct method of estimating the phase spectrum 
of the medium (averaged over positive and negative 
$\kappa_y$ values).  This is analogous to the 
reconstruction of an image from a hologram.

The final result (\ref{eq:pdi-weak3}) can be viewed as 
the interference of scattered waves with an unscattered
wave. To see this compare $S_2(f_{\lambda},\ft)$,
(excluding the $P_w$ term) with the interference 
result in Eq.~\ref{eq:fpqg1} discussed 
in \S\ref{sec:pt+ext}.
First we transform into scaled variables ($p,q$)
and assume small fractional differences in wavelength, for
which $f_{\lambda}\lambda_0 \sim f_{\nu} \nu_0$.
Hence we can express $\kappa_{yp}$ in
Eq.~\ref{eq:kyp} as:
\be
\kappa_{yp} = \sqrt{|p|-q^2} /(s\,s_0) \; ,
\ee
so $H(\kappa_{yp}^2)/\kappa_{yp}$ becomes
$H(U)/\sqrt{U} (s\, s_0 )$, where $U$ is defined 
as in the interference result for 
$S_2(f_{\lambda},\ft)$ (Eq.~\ref{eq:pdi-weak3}) 
with $\psivec_a=0$, and $s_0$ is the diffractive
scale as defined in \S\ref{sec:powerlaw-sfn}.  In 
Eq.~\ref{eq:fpqg1}, the   
brightness function $g$ represents the scattered waves
that interfere with an undeviated plane wave,
corresponding to the mean intensity in weak scintillation. 
%We note that $g$ is the angular spectrum corresponding to
%$\f$, which in weak scintillation is
%linearly related to the deviation in screen phase,
%which in turn has a wavenumber spectrum $P_{\phi}$.
%This gives a qualitative explanation of
%how $g$ maps into $P_{\phi}$.

\subsection{Extended Source}

A temporally and spatially incoherent extended source 
at a distance $D$ from the observer
is described by its brightness distribution 
$B_{\rm sou}(\thvecp)$. Hence we can simply add 
the intensity patterns 
due to each point component at $\thvecp$ to obtain
the well-known convolution result:
\be
I_{\rm ext}(\rvec) = \int d\thvecp \;
I(\rvec+\thvecp D(1-s),0,\lambda)  \; B_{\rm sou}(\thvecp) \; ,
\label{eq:iext}
\ee
where $I(\rvec,\rsvec,\lambda)$ is the intensity 
pattern for a point source at $\rsvec$. This convolution can also
be expressed in the wavenumber ($\kappa$) domain as a product
using the source visibility function 
$V(\bfu=\kappavec D(1-s)/2\pi)$, where $\bfu$
is the baseline scaled by the wavelength. 
Combining this relation with the point-source expressions, we find that
the integrand in Eq.~\ref{eq:pdi-weak2}
is multiplied by the product
$V_1(\bfu=\kappavec D(1-s)/2\pi) V_2^{*}(\bfu=\kappavec D(1-s)/2\pi)$,
where $V_1,V_2$ are visibilities at $\lambda_1,\lambda_2$.
Now consider the wavelength dependence of the visibility
function. If the source brightness distribution is independent
of wavelength (i.e.\ fixed angular size), then 
$V_1(\bfu) = V_2(\bfu)$.  Consequently, 
in Eq.~\ref{eq:pdi-weak3} $P_{\phi}$
is simply multiplied by $|V|^2$ to give:
\be
S_2(f_{\lambda},\ft) &=& 
\frac{8\pi^3}{\vperp \dse \kappa_{yp}} H(\kappa_{yp}^2)
\sum_{\pm}
|V(f_t D(1-s)/\vperp,\pm\kappa_{yp} D(1-s)/2\pi)|^2
P_{\phi}(\kappa_x=\frac{2\pi f_t}{\vperp},\pm\kappa_{yp}) \nonumber \\
&&\quad\quad + \delta(f_{\lambda}) P_{\rm w,ext}(f_t)  \; , 
\label{eq:pdi-weak4}
\ee
where the summation is over two equal and opposite values for
$\kappa_{\rm yp}$.
In this equation the $P_{\rm w,ext}$ function is similarly modified
by the visibility function but is of no immediate interest here.
Our discussion shows that the secondary spectrum,
%$P_{\Delta I}(\ft,f_{\lambda})$, 
$S_2(f_{\lambda}, \ft)$
in the weak scintillation regime
can be inverted to estimate the product of the medium
phase spectrum by the squared visibility function of the source --
in two dimensions. Since there are several lines of evidence supporting
a Kolmogorov model for the phase spectrum, 
we have a new way of estimating the squared visibilty function 
of a source. This allows a form of imaging from spectral 
observations with a single dish.

\end{document}